\documentclass[11pt,preprint]{aastex}
\usepackage{amsmath}

\def\deg{\hbox{$^\circ$}}

\slugcomment{Accepted by ApJ}
\shorttitle{The X-ray Jet in Cen A}
\shortauthors{Kataoka et al.}

\begin{document}

\title{
The X-ray Jet in Centaurus A:\\
Clues on the Jet Structure and Particle Acceleration}

\author{Jun Kataoka$^{1}$, \L ukasz Stawarz$^{2, \, 3}$,
Felix Aharonian$^2$, Fumio Takahara$^4$, \\
 Micha\l\hspace{1mm} Ostrowski$^3$,  and Philip G. Edwards$^5$}
\affil{$^1$ Tokyo Institute of Technology, Meguro, Tokyo, Japan}
\affil{$^2$ Max-Planck-Institut fur Kernphysik, Heidelberg, Germany}
\affil{$^3$ Obserwatorium Astronomiczne, Uniwersytet Jagiello\'nski, 
Krak\'ow, Poland}
\affil{$^4$ Osaka University, Toyonaka, Osaka, Japan}
\affil{$^5$ Institute of Space and Astronautical Science, JAXA,
Sagamihara, Japan}

\altaffiltext{1}{e-mails: kataoka@hp.phys.titech.ac.jp ; 
Lukasz.Stawarz@mpi-hd.mpg.de ; Felix.Aharonian@mpi-hd.mpg.de ; 
takahara@vega.ess.sci.osaka-u.ac.jp ; mio@oa.uj.edu.pl ; pge@vsop.isas.jaxa.jp}

\begin{abstract}
  We report detailed studies of the X-ray emission from the
  kiloparsec-scale jet in the nearest active galaxy, Centaurus~A. By
  analyzing the highest quality X-ray data obtained with the {\it Chandra}
  ACIS-S, 41 compact sources (mostly bright jet-knots) were found
  within the jet on angular scales less than 4~arcsec, 13 of which
  were newly identified. We construct the luminosity function for
  the detected jet-knots and argue that the remaining emission is
  most likely to be truly diffuse, rather than resulting from the sum
  of many unresolved fainter knots. 
We subtracted the contributions of the bright knots from the total
X-ray jet flux, and show that the remaining extended emission
has a relatively flat-topped intensity profile in
the transverse jet direction, with the intensity
peaking at the jet boundaries between 50$''$ and 170$''$.
  We note that limb-brightened morphologies have been observed previously at
  radio frequencies in a few FR~I and FR~II jet sources, but never so
  clearly at higher photon energies. Our result therefore supports  a
  stratified jet model, consisting of a relativistic outflow including
  a boundary layer with a velocity shear. In addition, we found that
  the X-ray spectrum of the diffuse component is almost uniform across
  and along the jet, with an X-ray energy spectral index of $\alpha_{\rm
    X} \approx 1$, similar to those observed in the compact
  knots. We discuss this spectral behavior within a framework of shock
  and stochastic particle acceleration processes, connected with
  the turbulent, supersonic, and non-steady nature of the relativistic
  outflow. We note some evidence for a possible spectral hardening 
  at the outer sheath of the jet, and manifesting itself in observed
  X-ray spectra of $\alpha_{\rm X} < 0.5$ in the most extreme cases. Due
  to the limited photon statistics of the present data, 
  further deep observations of Centaurus~A  
  are required to determine the reality of this finding, however we note that
  the existence of the hard X-ray features at outer jet boundaries
  would provide an important challenge 
  to theories for the evolution of
  ultra-relativistic particles within extragalactic jets.
\end{abstract}


\section{Introduction}

The {\it Chandra X-ray Observatory} has confirmed that X-ray
emission from large-scale jets is common in radio galaxies and quasars
(e.g., Harris \& Krawczynski\ 2002; Kataoka \& Stawarz\ 2005, and
references therein). In nearby FR~I sources, the typical
radio-to-X-ray spectrum of kpc-scale jet knots is consistent with a
single, smoothly broken power-law continuum, suggesting that this
broadband emission is entirely due to nonthermal synchrotron radiation
from a single electron population (e.g., Hardcastle et al.\ 2001 for
3C\,66B), although spatial offsets observed in these sources between
the positions of the X-ray and radio knots may indicate a more
complicated physical picture. Meanwhile, the observed X-ray knots in
hundred-kpc--scale quasar jets are much brighter than expected from a
simple extrapolation of the radio-to-optical fluxes (e.g., Schwartz et
al.\ 2000 for PKS\,0637$-$752). This has led to the hypothesis that
the X-ray production involved inverse-Comptonization of CMB photons
within a highly relativistic flow (Tavecchio et al.\ 2000, Celotti et
al.\ 2001). It has been argued, however, based on recent studies of
multi-wavelength properties of powerful large-scale jets, that the
inverse-Compton model may face serious problems (see, e.g., the
discussion in Stawarz et al.\ 2004; Kataoka \& Stawarz\ 2005).
Moreover, the inverse-Compton model cannot readily explain the
detected X-ray emission from a few FR~II jets --- Pictor~A (Wilson et
al.\ 2001), 3C\,303 (Kataoka et al.\ 2003), 3C\,120 (Harris et al.\
2004), 3C\,403 (Kraft et al.\ 2005) --- since they are 
expected to be
the $de$-$beamed$ analogues of the radio-loud quasars. Therefore, the
synchrotron origin of the keV photons remains a plausible alternative
even for powerful FR~II/quasar jets.

The synchrotron model for X-ray emission from the large-scale jets in radio
galaxies (FR~I/FR~II) and quasars involves much lower bulk velocities
for the flows, but on the other hand requires the presence of
ultra-relativistic electrons with energies $E_{\rm e} \ge 100$\,TeV. 
This requirement is supported in a natural way by the
expectation that the large scale jets are good candidates for the
the acceleration of cosmic ray particles owing to their large, 
extended structures and
turbulent nature (see Hillas\ 1984). From the theoretical perspective,
it has been proposed that the most energetic particles are accelerated
preferentially in the turbulent boundary layers of such outflows
(Ostrowski\ 2000; Stawarz \& Ostrowski\ 2002; see also in this context
Rieger \& Duffy\ 2004 and references therein). In fact, the
`limb-brightened' jet morphology expected in this type of models has
been observed at radio frequencies (probing relatively low electron
energies) in several objects, such as M\,87 (FR~I, Owen et al.\ 1989)
and 3C\,353 (FR~II, Swain et al.\ 1998). More importantly, the continuous
acceleration processes acting thereby should result in the formation of
flat spectra at the highest electron energies, which may account for the
hard X-ray spectra and high X-ray luminosities of powerful jet
sources. Similarly, Dermer \& Atoyan (2002) suggested the formation of
high-energy spectral hardening owing to inefficient Klein-Nishina
radiative losses of electrons, which also could appear at X-ray
energies for some particular choices of jet parameters. Therefore
the investigations of (1) the transverse structure of X-ray jets, 
and (2) the spectral properties of diffuse X-ray jet
emission, provide important clues for an understanding of the
production of cosmic ray particles and the origin of hard X-ray
emission in large scale jets. Unfortunately, only few such studies 
have been reported so far, mainly due to the technical
difficulties. In fact, it is generally impossible to resolve jets in
low power FR~I radio galaxies or distant quasars, even with the
excellent spatial resolution (0.5$''$ half power diameter on-axis) and
sensitivity of {\it Chandra} (see {\it Proposers' Observatory Guide v.7}\,\footnote{\texttt{http://cxc.harvard.edu/proposer/POG/index.html}}).

Centaurus~A (hereafter Cen~A) is the nearest active galaxy
($d = 3.4$\,Mpc, for which 1$''$ corresponds to 16\,pc) and has been well
studied across the entire electromagnetic spectrum at the highest
linear spatial resolution (Israel\ 1998). It is considered to be 
a prototypical FR~I radio galaxy, although the one-sidedness of its kpc-scale 
jet and the presence of the outer, Mpc-scale radio lobes is unusual 
for this type of object. Previous observations of
Cen~A have resulted in the detection of a bright X-ray and radio jet
extending $\ge$\,4$'$ northeast of the nucleus (Schreier et al.\ 1979,
1981; Clarke et al.\ 1986, 1992; Kraft et al.\ 2002; Hardcastle et
al.\ 2003). From the analysis of the broadband radio--to--X-ray jet
emission, the synchrotron origin of keV photons has been established
for the bright knots, which are characterized by a minimum pressure
$\sim 10^{-9}$\,dyn\,cm$^{-2}$, equipartition magnetic field $\sim
100$\,$\mu$G and $1-10$\,keV luminosities of the order of $\sim
10^{39}$\,erg\,s$^{-1}$. In addition, a very complex morphology of the
whole outflow was noted in both the radio and X-ray bands, including the
presence of bright filaments and diffuse extended regions with complex
sub-structure. The diffuse X-ray emission clearly extends to the edge
of the radio jet and is reasonably well matched to the radio emission on
these scales. 
Moreover, Hardcastle et al.\ (2003) noted two
conspicuous regions where the extended X-ray emission appeared to be
associated with the edges of the jet, and to be absent in the center.
These X-ray edges of the jet have no obvious radio counterparts,
although Clarke et al.\ (1986, 1992) also reported a limb-brightened
radio morphology in some sections of the jet. We note that these radio
polarization studies indicated a projected magnetic field parallel to the
jet axis along the whole jet length.

Stimulated by these previous reports, the goal of this work is
twofold, namely determining the transverse profile of the diffuse
X-ray emission extending over the Cen~A jet, and constraining its
spectral properties. Our approach differs from previous studies in
that we focus on the diffuse emission rather than the bright jet
knots, which are thought to be compact shocks inside the jet
(Hardcastle et al.\ 2003). We remove the contributions from the knots,
as particle acceleration may take place in quite
different manners in different jet regions (but we will revisit this
problem later in the discussion section). The paper is organized as
follows: In $\S$2, we describe the {\it Chandra} observations, data
reduction process and our analysis method. In $\S$3, we
present the results of the data analysis. In $\S$4, we
discuss our findings in the context of particle acceleration models,
presenting our conclusions in $\S$5.

\section{Data Reduction}

Since the launch of {\it Chandra} X-ray Observatory in July 1999,
X-ray observations of Cen~A have been conducted 10 times, three of
which were for calibration (CAL) purposes, and the remaining for
General Observer (GO) or Guaranteed Time Observing (GTO) proposals.
The first {\it Chandra} X-ray image of Cen~A was obtained with the
microchannel plate X-ray imaging detector HRC-I (Obs ID 463, 1253, and
1412), and the results presented in Kraft et al.\ (2000). The first
useful spectral information was subsequently presented based on
observations using the ACIS-I configuration (Obs ID 316 and 962: Kraft
et al.\ 2002). The exposure times for the two observations were
35.9\,ksec and 36.5\,ksec, respectively. Although the jet was 4$-$5$'$ 
off-axis from the aim point in these observations, and
the image resolution was therefore significantly degraded, 31 knots in
the Cen~A jet were successfully detected.

More recent {\it Chandra} observations of Cen~A were made in 2002
September (Obs ID 2978) and 2003 September (Obs ID 3965) using the 
standard ACIS-S configuration, where the back-illuminated S3 chip was
focused to obtain maximal sensitivity to soft photons. The exposure
times for the two observations were 45.2\,ksec and 50.2\,ksec,
respectively. The active nucleus was at the aim point, resulting in
both sub-arcsecond resolution and good spectral sensitivity for the
whole inner part of the jet in Cen~A. Although the imaging degrades
substantially for sources more than 1$'$ off-axis, and the X-ray jet
in Cen~A extends over $\sim$4$'$ in length, the encircled energy
for a point source is contained within a diameter of $\sim$3$''$ all along the
outflow. The roll angle of the satellite was carefully chosen to avoid
the readout streak of the bright nucleus impinging on the jet image.
The results from the first observation, Obs ID 2978, are
given in Hardcastle et al.\ (2003). The X-ray image was compared with
new, high dynamic range VLA images for one-to-one identification of
radio knots, for which apparent sub-luminal motions were found (on
scales of hundreds of pc). Here we analyze the combined ACIS-S data
(Obs ID 2978 and 3965) to obtain the best image resolution and
spectral information reported to date.
 
A co-added, exposure-corrected image of the Cen~A jet in the
0.4$-$8\,keV bandpass is shown in Figure~1 ({\it upper}). The image is
smoothed with a $\sigma$\,=\,0.5\,arcsec Gaussian function. The
nucleus is located in the southwest corner of the image, and the jet
extends to the northeast. The figure shows that the diffuse emission
from the jet is observed up to $\sim$ 4$'$ from the nucleus. Also, the
{\it Chandra} image shows that the X-ray jet consists of many knots of
enhanced emission embedded within the diffuse regions, as described in
detail by Kraft et al.\ (2002) and Hardcastle et al.\ (2003). Since
the primary purpose of this paper is to study the characteristics of
the {\it diffuse} X-ray emission associated with the jet, our approach
is complementary to previous reports in the literature: we first resolve 
the point-like sources within the jet (in projection).
\footnote{From 
a comparison of the source density within the jet and that
in surrounding regions, it is highly likely that most of sources
are jet knots, with only a few likely to be background, or
foreground, sources (mainly X-ray binaries in Cen A).  
A check of the NASA/IPAC Extragalactic Database (NED) yielded only 
one possible identification, with an X-ray star. Similarly 
the chance probability that one of the 
new X-ray point sources discussed here 
is a background AGN is less than 1\% (Kraft et
al.\ 2001). See Table~1 for more details.}
Next we subtract the contributions from these point sources 
to enable a quantitative measurement of the
diffuse X-ray emission. One may ask if the remaining emission is
truly diffuse in nature, or whether it is the summation of numerous
unresolved jet-knots at even smaller spatial scales. We will revisit
this problem in $\S$3.3, based on the luminosity function of
the jet-knots.

Using a wavelet decomposition source-detection algorithm 
({\it wavdetect\/}; Vikhlinin et al.\ 1995) to detect emission enhancements
on angular scales of 0.5$''$ and 1$''$, we find a total of 35 distinct
knots or enhancements of X-ray emission embedded within the diffuse
regions. We then extend the angular scales to 2$''$ and 4$''$, but
only an additional six sources are found in these larger region sizes.
This suggests that most of the jet knots in Cen~A are very compact and
that their extent is smaller than, or comparable to, that expected
from the {\it Chandra} point spread function. Our result is therefore
consistent with findings by Kraft et al.\ (2000) and Hardcastle et
al.\ (2003), who claimed that most of the X-ray knots are diffuse but
at the 1$''$ ($\sim$\,10\,pc) level. Due to the good photon statistics
and excellent angular resolution during the observations, the total
number of the detected sources is significantly larger than that
reported in the literature (i.e., 41 compared with 31 for Kraft et al.\ 2002).
Three of the weaker knots reported in Kraft et al.\ (2002), NX1 (Knot-1),
EX3 (Knot-19), and FX4 (Knot-23), were not detected at the previous
level in our high resolution, co-added ACIS-S image.
The location, distance and position angle of the 13
knots or enhancements which were newly detected in our analysis
are listed in Table~1.

We next apply {\it dmfilth} implemented in the CIAO 3.2 package to
replace the pixel values for the point sources detected by 
{\it  wavdetect} with values interpolated from surrounding background
regions (DIST option in {\it dmfilth}; see 
{\it Science Threads for 
CIAO 3.2}\,\footnote{\texttt{http://cxc.harvard.edu/ciao/threads/index.html}}).
Here the source regions are defined as the ellipses output from 
{\it wavdetect} which contain at least a 3$\sigma$ excess of
photon counts over the background. The background is taken from the 
``doughnut'' shaped region
surrounding the source, with the dimensions of the major and minor
axes of the outer ellipse both doubled. The resultant image is almost
flat, as expected, but contains a few irregular pixels with
outstanding photon counts. This is because, in some regions, source
ellipses overlap. The background region for a certain point source
therefore inevitably contains photons from neighboring source regions.
We therefore repeat the same process (i.e., source detection by {\it
  wavdetect} and smoothing by {\it dmfilth}) three times, until no
irregular bins appeared in the image. The resultant X-ray image of the
diffuse emission is given in Figure~1 ({\it bottom}: 0.4$-$8.0\,keV).
We also made bandpass images for soft X-rays (0.4$-$1.5\,keV) and
hard X-rays (1.5$-$8.0\,keV) using the same procedure.

\section{Data Analysis and Results}

In order to derive the intensity (photon counts) profile of the jet 
both in the longitudinal and transverse directions, we first define 
the position angle ($\Theta$) and the distance ($r$) with respect to 
the nucleus of Cen~A as shown in Figure~1 ({\it bottom}). 
Note that the position angle is measured from east through north 
(i.e., in a clockwise direction in the Figure). 
Then we can integrate the photon counts in a sector of an arbitrary 
range of $r$ and $\Theta$, to construct the
intensity profile parallel or perpendicular to the jet main axis,
which is given by $\Theta \sim 35\deg$. Since our approach is simply 
to sum the counts in each sector bounded by the stated values of $r$ 
and $\theta$, and not to derive the emissivity, no normalization by 
the area of each sector ($\propto$ $1/r^2$) was performed.

\subsection{Longitudinal Jet Profiles}

We first integrate the photon counts in each thin shell within a
radial distance $r$ $\sim$ $r$+$\Delta r$ over the range of position
angle 25$\deg$ $\le$ $\Theta$ $\le$ 45$\deg$, and then construct the
longitudinal intensity profile in $\Delta r$\,=\,4$''$ steps, from
$r$\,=\,0$''$ to 250$''$. Figure~2 shows the longitudinal jet profiles
thus produced, for both soft X-rays (0.4$-$1.5\,keV: {\it top})
and hard X-rays (1.5$-$8.0\,keV: {\it bottom}). The
dotted histograms show the profiles derived from the raw X-ray images,
whereas those for the diffuse X-ray emission are given as the data
points on the solid-line histogram. We also show the intensity profile
of the background (i.e., off-jet) region as a dashed line for 
comparison, where the background was accumulated in the 
15$\deg \le \Theta \le 25\deg$ and 45$\deg \le \Theta \le 55\deg$ regions. 
The intensity profiles along the jet are quite
different in the two energy bands, especially at the innermost part of
the jet (0$\le r \le 50''$), where the soft X-ray photons are
significantly absorbed in the dust lane crossing the Cen~A host
galaxy.

To see this more clearly, we divided the 1.5$-$8.0\,keV longitudinal
jet profile by that of the 0.4$-$1.5\,keV, to check for hardness
variations along the X-ray jet. Figure~3 compares the hardness ratio
(HR) and the 0.4$-$8.0\,keV X-ray intensity profile. One can clearly
see that HR changes significantly at $r_{\rm c} \simeq 50''$,
corresponding to a linear scale of $\sim$ 800\,pc.
Obviously, such a dramatic change of the HR is not due to
the changes of intrinsic jet spectrum. In fact, we can see that the HR
changes in a similar way for the background region also, as shown
by the dashed line in the bottom panel. The inner 1\,kpc of the jet
is observed through the dust lane of the galaxy, making detection of any
optical component difficult (Israel\ 1998). This same dust lane, or
equatorial gas, heavily absorbs soft X-rays as well, resulting in
the significant changes of the hardness ratio seen in Figure~3. In the
outer part of the jet (50$'' \le r \le 240''$), the HR is
almost constant. This trend suggests that the X-ray spectrum does not
change significantly along the jet, but the contribution from the
background photons must be carefully examined before reaching this
conclusion.
 
To find possible spectral changes along the jet, we fit the diffuse
emission with a power-law function in XSPEC, for each shell in $\Delta
r$\,=\,10$''$ steps between 10$''$ and 100$''$, 
and 20$''$ steps between 100$''$
and 240$''$. We do not examine the innermost part of the
jet ($r \le 10''$) because the X-ray emission from the nucleus
significantly contaminates the jet emission there.\,\footnote{The 
contamination level is quite uncertain, as the bright nucleus 
emission is strongly piled up. We however make a crude estimate that 
more than 80\% of the photons come from the nucleus, rather than from the 
jet,  within the most inner part ($r$ $\le$ 10$''$) of the jet. See
Fig.~2  for a comparison of background profile with the diffuse jet emission 
profile.} 
We take the background from neighboring shells just outside the jet
(15$\deg$\,$\le$\,$\Theta$\,$\le$\,25$\deg$,
45$\deg$\,$\le$\,$\Theta$\,$\le$\,55$\deg$ at the same range of $r$)
and subtract it from the source spectrum. The result of spectral
fitting along the jet is shown in Figure~4. As expected from the
hardness profile, the X-ray spectrum clearly changes at a
characteristic distance of $r_{\rm c}$ $\simeq$ 50$''$. In the inner
jet region ($r \le r_{\rm c}$), the spectrum is well represented
by a power-law function heavily absorbed with a free zero-redshift
absorbing column. A combined spectrum between $r=10''$ and 50$''$
gives the best-fit absorbing column density of $N_{\rm H}$ =
(4.79$^{+0.68}_{-0.53}$)$\times$10$^{21}$\,cm$^{-2}$ and 
an intrinsic energy
spectral index of $\alpha_{\rm X}$ = 0.89$^{+0.17}_{-0.13}$, where the
reduced $\chi^2$ is 0.50 for 164 degrees of freedom (dof).

The absorption column in the outer region ($r$ $\ge$ $r_{\rm c}$) is
much smaller, and similar to
the Galactic column density
($\sim$0.7$\times$10$^{21}$\,cm$^{-2}$). Although there is a weak
trend of spectral steepening at larger distances from the nucleus,
there is no sudden change in the spectral form over the wide distance
of $r=50''$ and 240$''$. The subtle anti-correlation between the
fitted X-ray spectral index and the X-ray flux which can be seen in
Figure~4 is most likely due to minor variation of the fitted absorbing
column density at $r \ge r_{\rm c}$. The best-fit parameters for
the average spectrum in this outer jet region are $N_{\rm H}$ =
(0.96$\pm$0.13)$\times$10$^{21}$\,cm$^{-2}$ and $\alpha_{\rm X}$ =
1.19$\pm$0.07, which have a reduced $\chi^2$ of 0.88 for 267 dof.
This energy spectral index of the diffuse emission inside the jet is
consistent with that reported in Hardcastle et al.\ (2003),
viz, $N_{\rm H}$ = (1.1$\pm$0.3)$\times$10$^{21}$\,cm$^{-2}$
and $\alpha_{\rm X}$ = 1.00$^{+0.16}_{-0.15}$, derived for the
extended jet emission in the jet between knot A2 and the knot B
region (approximately 20$''$ $\le$ $r$ $\le$ 70$''$), 
excluding all compact X-ray features. Interestingly, in the
100\,kpc-long jet in the radio galaxy Pictor~A, the observed X-ray
spectral index is constant along the jet, and is $\alpha_{\rm X}$
$\sim$ 1 (Wilson et al.\ 2001), as in the kpc-scale jet analyzed here,
suggesting the action of similar particle acceleration processes for
the high-energy electrons resulting in the synchrotron X-ray emission.

\subsection{Transverse Jet Profiles}

We next integrate the photon counts in each narrow sector with an
angular width $\Delta \Theta$ over certain radial distances (selected
in the range 50$''$\,$\le$\,$r$\,$\le$\,250$''$), and then construct the
transverse intensity profiles in $\Delta \Theta$\,=\,1$\deg$ steps,
from $\Theta$\,=\,20$\deg$ to 50$\deg$. We do not examine the
innermost part of the jet ($r$\,$\le$\,50$''$) because of the strong
absorption due to the dust lane, as discussed in the previous section.
Moreover, we cannot determine the position angle $\Theta$ in the
inner-jet region with sufficient accuracy. For example, the
$\sim$\,1$''$ diameter of a typical {\it Chandra} image resolution
corresponds to $\Delta \Theta$\,$\ge$\,1$\deg$ at $r$\,$\le$\,50$''$,
which is larger than the angular step size of $\Delta
\Theta$\,=\,1$\deg$ for our analysis. Note that $\Delta
\Theta$\,=\,1$\deg$ at $r$\,=\,250$''$ corresponds to
$\approx$\,4.4$''$, or to a linear scale of $\approx 70$\,pc. Figure~5
shows the transverse jet profiles thus produced for both soft
X-rays (0.4$-$1.5\,keV: {\it top}) and hard X-rays
(1.5$-$8.0\,keV: {\it bottom}). We made separate plots of the
transverse profiles for $r \le 170''$ and $r\ge170''$, because
the jet's emission decreases to the background level at 170$''$, but then
gradually increases again to a peak at $r \sim 210''$
(see Figures~1 and 2). The dotted histograms show the profiles derived
from the raw X-ray images, whereas those for the diffuse X-ray
emission are given as the data points on the solid-line histogram.
The transverse jet profile of the diffuse emission shows a flat-top 
structure for the jet region between 50$''$ and 
170$''$ but, in contrast, a single, narrow peak for
the outermost part of the jet (170$''$ $\le$ $r$ $\le$ 250$''$).  Such an
edge-brightened structure, as observed for the inner jet region, is
not readily apparent in the asymmetric intensity profile derived from
the raw data.

The flat-topped intensity profile along the transverse jet
direction in Cen~A cannot be explained by a homogeneous plasma fluid
moving with a uniform velocity. One natural and simple possibility
to account for the observed profile is to consider a stratified model
for the kpc-scale Cen~A jet, with a velocity shear across the jet
outflow. In general, such a structure is actually required to explain
many observed properties of extragalactic jets, and is also expected
on the grounds of theoretical and numerical jet studies (see the discussion
in Stawarz \& Ostrowski\ 2002 and references therein). In this model,
different Doppler enhancements of the emission produced within
different velocity components can result in the observed
limb-brightening of the jet. The other possibility discussed in the
literature is to consider a mildly-relativistic jet bulk velocity and
the enhanced emissivity of the non-thermal plasma at the jet
boundaries resulting from, e.g., non-uniform radial structure of the
jet magnetic field due to jet interaction with surrounding medium or
some particular large-scale magnetic topology (see, e.g., the
discussion in Owen et al.\ 1989). Because of the apparent
one-sidedness of the kpc-scale jet in Cen~A, we favor the former
possibility as the dominant factor shaping the transverse intensity
profile, although both Doppler favoritism and plasma emissivity
stratification may both play a role.

The exact velocity profile across the kpc-scale jet in Cen~A cannot be
precisely modeled with the present data. Let us however underline a
few appropriate constraints. First, note that the observed parallel
magnetic field structure along the jet axis, with no evidence for the
transverse component that is usually observed in FR~I sources, is
consistent with a well-developed velocity shear (see in this context
Laing \& Bridle\ 2002). Second, if the limb-brightening of the diffuse
X-ray emission is predominantly due to varying Doppler enhancements of
different jet axisymmetric layers, and the emissivity is $uniform$ along 
the jet (although this may not be a realistic assumption for a stratified jet),
a component for which the emission is most strongly beamed toward 
the observer is the one characterized by the bulk velocity
$\beta(\Theta) = \cos \alpha$ (or bulk Lorentz
factor $\Gamma (\Theta) = 1 / \sin \alpha$), where $\alpha$ is the jet
viewing angle.  With the preferred $\alpha \sim 50 \deg$ 
(Tingay et al.\ 1998; Kraft et al.\ 2003), 
one can roughly estimate the bulk velocity
of the limb-brightened regions ($\Theta \sim 30 \deg$ and $\sim 40
\deg$) as $\beta \sim 0.65$ (or $\Gamma \sim 1.3$). 
This very rough, illustrative estimate indicates that the 
spine of the kpc-scale jet in Cen~A --- at the position angle $\Theta =
35 \deg$ --- is most likely still relativistic, with a bulk Lorentz
factor $\Gamma \geq 2$. 
In this context, the single narrow peak at the outermost part of the jet
(170$''$ $\le$ $r$ $\le$ 250$''$), may correspond to the spine,
decelerated significantly at the terminal part of the collimated
outflow. Finally we note that the half-opening angle of the X-ray jet,
$\phi_{\rm X} \leq 5 \deg$, is smaller than that  of the
radio jet structure $\phi_{\rm R} \sim 10 \deg$ for $r > 60''$ (Clarke
et al.\ 1992; see also Kraft et al.\ 2002). 
In the framework of the above discussion, this would be
consistent with the boundary shear layer spreading in the transverse
direction as the jet propagates.

Similar to the analysis in the longitudinal direction, we divided the
1.5$-$8.0\,keV transverse jet profile by that of the 0.4$-$1.5\,keV,
to examine the hardness ratio (HR) variation across the X-ray jet.
Figure~6 compares the HR and the 0.4$-$8.0\,keV X-ray intensity
profiles. Here we divided the inner jet region further into three
separate sectors, (a) 50$''$ $\le$ $r$ $\le$ 80$''$, 
(b) 80$''$ $\le$ $r$ $\le$ 120$''$, and (c) 120$''$ $\le$ $r$ $\le$
170$''$, for a detailed
comparison of changes in transverse profile. Edge-brightened structure
is most clearly seen in the profile derived for
50$''$ $\le$ $r$ $\le$ 80$''$ (Figure~6a). 
Interestingly, one can clearly see that
the HR is almost uniform across the jet, but shows a weak sign of
enhancement at the very edges of the X-ray jet regions ($\Theta$
$\sim$ 25$\deg$ and $\Theta$ $\sim$ 45$\deg$), followed by a sudden
``drop'' (softening) close to the edges. Similar hardening or
softening may explain some variation of the HR even for the jet-edges in
Figure~6 (b) and (c). 
We may speculate that
the spectrum changes considerably at the outer edges of the jet, 
accounting for these variations, 
though the present data do not allow this to be established.

First, since the HR contains a significant contribution from
background photons, apparent changes in HR might be due to 
variation in the background characteristics rather than changes in the
source spectrum itself. This effect is important especially at the jet
edges, where the emission is dominated by background photons. We
will carefully examine this problem later by choosing an appropriate
background for spectral fitting. Second, it is well known that the
response of the {\it Chandra} X-ray mirror changes with the incident
X-ray energy. The half-power diameter of the point spread function
is 0.5$''$ at 1\,keV, but slightly worse at higher energies (e.g.,
1$''$ at 5\,keV; see {\it The Proposers' Observatory Guide\/}.
Such an energy dependence of the angular resolution might introduce HR
variations perpendicular to the Cen~A jet, especially at the jet
edges. To check for this effect quantitatively, we smoothed the
0.4$-$8.0\,keV image (Figure~1 {\it bottom}) with Gaussian functions
of $\sigma$ = 0.5$''$ and 1.0$''$ to mimic the images obtained in
different energy bands. The resultant intensity profiles have only
negligible (less than 2$\%$) changes in the HR across the jet,
suggesting that we cannot explain the observed hard X-ray features
shown in Figure~6 by this effect alone.

Finally, variations of HR may be caused by any possible artifact
caused by {\it dmfilth} and inclusion of background.  We therefore
remove all the point sources detected by {\it wavdetect} and fit the
diffuse emission with a power-law function in XSPEC, without filling
the `holes' remaining after subtracting the source ellipses. We assume
a fixed absorption column density of $N_{\rm H}$ =
0.96$\times$10$^{21}$\,cm$^{-2}$, which is valid all along the jet if
$r$ $\ge$ 50$''$ (see $\S$3.1). We take the background from two
sectors just outside the jet
(19$\deg$\,$\le$\,$\Theta$\,$\le$\,23$\deg$,
47$\deg$\,$\le$\,$\Theta$\, $\le$\,51$\deg$ within an equal $r$ $\sim$
$r$+$\Delta$$r$) and subtracted them from the source spectrum. The
result of spectral fitting across the jet with $\Delta
\Theta$\,=\,2$\deg$ steps is shown in Figure~7. Although the photon
counts are inevitably poor at the jet edges, the general trends of the
flux variations and spectral changes are very similar to that expected
from the HR profiles. The spectral index seems to increase slightly and
gradually along the main jet, being $\alpha_{\rm X}$ $\ge$ 1.0. In
addition, the hardest X-ray spectra are observed at the edges of the jet
in the 50$''$ $\le$ $r$ $\le$ 80$''$ region, where $\alpha_{\rm X}$ =
0.23$^{+0.93}_{-0.71}$ for $\Theta$$\sim$25$\deg$ or $\alpha_{\rm X}$
= 0.40$^{+0.64}_{-0.58}$ for $\Theta$$\sim$45$\deg$. We note that the
background spectrum is approximately represented by a power-law with
an energy spectral index of $\alpha_{\rm X}$ $\sim$ 1.5$-$2.0, and
therefore background contamination cannot produce an
X-ray spectrum as hard as that as observed in Figure 7(a). 
Unfortunately, the photon
statistics are insufficient to confirm that the X-ray spectrum
is actually harder at the very edges of the jet. Nevertheless, this
provides an important motivation for future deep observations of the
Cen~A jet.
 
\subsection{Diffuse or Unresolved?}

As pointed out by Kraft et al.\ (2002), the X-ray morphology of
the Cen~A jet is much more complex when observed at higher spatial
resolution. This raises the question whether some fraction of the extended
emission discussed in this paper can be explained by the pile-up of small
scale knots below the resolution limit for {\it Chandra} point
source detection. To estimate what part of the remaining unresolved
emission of the jet could indeed result from a large number of
unresolved low luminosity compact features, we construct a luminosity
function (LF) for the already resolved jet-knots, presented in
Figure~8. Here we consider the 32 jet-knots located between
50$''$ and 250$''$ as the strong absorption due to
the dust lane makes determination of the unabsorbed luminosity
quite uncertain for $r$ $\le$ 50$''$. We assume a power-law energy spectrum
with the spectral index $\alpha_{\rm X}$ = 1.0 (Hardcastle et al.\
2003; Kraft et al.\ 2002) and $N_{\rm H}$ =
0.96$\times$10$^{21}$\,cm$^{-2}$ (see $\S$3.1) to estimate the 
0.5--5\,keV luminosities of the knots. 
Although Kraft et al.\ (2002) derived their LF
for 0.1--10\,keV luminosities of the knots with a different choice of
spectral parameters ($\Gamma_{\rm X}$ = 2.3 and $N_{\rm H}$ =
1.7$\times$10$^{21}$\,cm$^{-2}$), a rough conversion to their 0.1--10\,keV 
fluxes can be obtained by multiplying our 0.5--5\,keV luminosities
by a factor of three.

The limiting sensitivity for this analysis is somewhat uncertain due to
source confusion and possible variations of spatial extent among the
jet-knots. Furthermore, the sensitivity of the observation varies across
the field of view due to the telescope vignetting and variations in the
PSF. A detailed computation of the limiting sensitivity requires a Monte
Carlo simulation taking into account the unknown morphology of the
\emph{intrinsic} diffuse emission and is beyond the scope of this
paper. Here we follow the simplified method adopted by Kraft et
al.\ (2001): we use the measured background and the telescope PSF at the
edge of the field of view and we require a 4\,$\sigma$ measurement of
the count rate. Our conservative estimate (using the background map and
the telescope vignetting; Kraft et al.\ 2001) provides a sensitivity
limit of $L_{\rm sens}$ $\sim$ 4$\times$10$^{36}$\,erg s$^{-1}$. Let us
briefly discuss this value. During these observations 
the encircled energy of a point source is contained within
a diameter of $\sim$3$''$ all along the jet. Assuming that the
intrinsic spatial extent of the knots is less than 4$''$, we
conservatively expect that all the compact sources can be detected
within a circle of $\sim$ 6$''$ diameter, even if they are in the
outermost part of the jet (i.e., $\ge$ 4$'$ off axis with respect to the
mirror focus). The typical background within this region is $\sim$15~cts
from the observational data. To detect a knot at the required
statistical level, we therefore need $\sim$25 net source counts
excluding the background photons. This corresponds to an X-ray
luminosity of $L_{\rm sens}$ = 2$\times$10$^{36}$\,erg s$^{-1}$. If a
similar knot is embedded in extended emission (as observed in
50$''$ $\le$ $r$ $\le$ 80$''$), $\sim$50 net photons are required over
the $\sim$120 background photons. It is therefore reasonable to assume
$L_{\rm sens}$ = 4$\times$10$^{36}$\,erg s$^{-1}$ in the following
discussion.

In Figure~8, the hatched region shows  
where our
sampling is incomplete and somewhat biased. The overall LF is well
modeled by a broken power-law, with the power-law slope
flatter below the break. The dashed line on Figure~8 shows the fit
considered hereafter, where the power-law slope ($\kappa$ of $N(>L_{\rm
X})$ $\propto$ $L_{\rm X}^{-\kappa}$) flattens significantly from
$\kappa$ = 1.4 to 0.5 below $L_{\rm brk}$ $\simeq$
9$\times$10$^{36}$\,erg s$^{-1}$. The maximum likelihood method of
Crawford, Jauncey \& Murdoch (1970), using separate data sets
below and above the break, supports this fit with an uncertainty (defined
here as a standard deviation) of $\sigma$\,$\simeq$\,0.4 for each
$\kappa$. However, due to the limited number of the detected jet-knots (24
for $L_{\rm X}$ $\ge$ 4$\times$10$^{36}$\,erg s$^{-1}$), the
Kolmogorov-Smirnov test (Press et al.\ 1992) indicates that a single
power-law model with $\kappa$\,$\sim$\,0.7 is also statistically
acceptable. Even so, a single steep power-law with $\kappa$ $\ge$
1.0 can be rejected at the $>$90\% confidence level. Assuming
further that the distribution of knot luminosities extends down to some
minimum value $L_{\rm min}$, the integrated total X-ray luminosity can
be evaluated quantitatively for both the resolved and the unresolved
jet-knots (see Kraft et al.\ 2002 for the detailed formalism). We note that
the total luminosity of the Cen~A jet (i.e., the sum of the jet-knots
and the unresolved diffuse emission) is 1.6$\times$10$^{39}$\,erg
s$^{-1}$. Meanwhile, by integrating the LF down to any value of $L_{\rm
min}$ we get $<$ 8.1$\times$10$^{38}$\,erg s$^{-1}$, at most, if we
assume the ``best-fit'' broken power-law LF described above. Note
that this is only 25$\%$ larger than the sum of the \emph{already
resolved} jet-knots. The situation is almost unchanged if we take a
single power-law function with $\kappa$\,$\simeq$\,0.7 for the LF. We
therefore conclude that about half of the total jet emission, $L_{\rm
X}$ $\sim$ 8$\times$10$^{38}$\,erg s$^{-1}$, is truly 
diffuse in nature.

We finally consider an apparent discrepancy between our results and
previous reports.
Kraft et al.\ (2002) suggested that the
diffuse X-ray emission of the Cen~A jet may be entirely explained by 
unresolved knots only if 
$L_{\rm min}$ = 3$\times$ 10$^{35}$\,erg\,s$^{-1}$ 
and there is no break in the LF below the sensitivity limit
of $L_{\rm sens}$ $\sim$3$\times$10$^{37}$\,erg s$^{-1}$ (corresponding
to $L_{\rm X}$ $\sim$1$\times$10$^{37}$\,erg s$^{-1}$ in our LF). There
are several possibilities to account for this discrepancy. First, the
sensitivity limit adopted in Kraft et al.\ (2002) is calculated for a
single $\sim$35\,ksec observation. As discussed in detailed in Kraft et
al.\ (2001), the sensitivity is limited by the source luminosity, not by
the background, over the most of the field of view. Therefore, our
analysis based on the co-added ACIS-S image (95\,ksec exposure in total)
provides a factor of $\sim$\,$\sqrt{3}$ deeper sensitivity for the source
detection. Moreover, the PSF in the inner jet is a factor of $\sim$2
smaller than previous observations (see $\S$2), which further improves 
the signal-to-noise ratio.  Second, by applying the same sensitivity 
limit as Kraft et al.\ (2002), it would be difficult to see a flattening in the
low-luminosity part of the LF. Thus, a steep power-law LF is consistent
with the work of Kraft et al.\ (2002; who obtained $\kappa$=1.1), although is
implausible when the low-luminosity range ($L_{\rm X}$ $<$
10$^{37}$\,erg s$^{-1}$) is taken into account. Third, it seems that
quite different conclusions can be reached for different choices of
$\kappa$. In fact, integration of the LF down to arbitrarily small
$L_{\rm min}$ approaches infinity if $\kappa$ $\ge$ 1. Future deep
observations of the Cen~A jet are therefore strongly encouraged, although
it seems to us reasonable that a significant fraction of the X-ray jet
emission in this source is actually diffuse.

\section{Discussion}

\subsection{The Jet}

The observed X-ray spectral index of the diffuse emission within the
whole jet --- including both the central flow and the
limb-brightened edges --- is steep, $\alpha_{\rm X}$ $\simeq$
1.0--1.2. The implied energy distribution of the X-ray synchrotron
emitting electrons therefore has to be $n_{\rm e}(\gamma) \propto
\gamma^{-s}$ with $s = 2 \, \alpha_{\rm X} + 1$\,=\,3.0--3.4, where
$\gamma$ is the electron Lorentz factor. Such a spectrum is very
similar to the one observed in the compact knots, which are believed
to be the sites of shock waves, suggesting that the same, or a similar,
particle acceleration process is acting in both regions. In the
case of the shock acceleration, however, one expects the energy
spectral index of the freshly accelerated electrons to be 
$s_{\rm in}$\,=\,2.0 in the strong non-relativistic regime, or 
$s_{\rm in}$\,$\gtrsim$\,2.0 when relativistic plasma velocities 
are considered (see, e.g., Kirk \& Duffy\ 1999, Ostrowski\ 2002). 
Thus, if first order Fermi
shock acceleration is indeed responsible for formation of the X-ray
(synchrotron) emitting electrons within the Cen~A jet, the observed
particle energy spectrum --- both within and outside the resolved
knots --- has to be already modified by the (dominant) synchrotron
losses, i.e., the spectral index of freshly accelerated high-energy
electrons has to be increased by the radiative cooling
$|\dot{\gamma}|_{\rm syn} =\sigma_{\rm T} \, \gamma^2 \, B^2 / 6 \pi
\, m_{\rm e} c$ to $s = s_{\rm in} + 1$ (the `strong cooling regime' with
homogeneous magnetic field $B$; see Kardashev\ 1962). In fact, the
appropriate cooling timescale is very short, $t_{\rm syn} = \gamma /
|\dot{\gamma}|_{\rm syn} \sim 20 \, B_{-4}^{-3/2} \,
\varepsilon_{10}^{-1/2}$\,yrs (for a jet magnetic field $B_{-4} \equiv
B / 10^{-4}$\,G $\sim$ 1, an observed synchrotron photon energy
$\varepsilon_{10} \equiv \varepsilon / 10$\,keV $\sim$ 1, and a jet
Doppler factor $\delta \sim 1$), consistent with this
interpretation. We note in this context that the detailed analysis of the
radio emission from a number of FR~I sources of Young et al.\
(2005) revealed a `universal' injection spectral index for the radiating
electrons of $s_{\rm in}$ = 2.1, in accordance with the above discussion.
We also note that synchrotron cooling within a non-uniform magnetic
field can result in a spectral break larger than the standard one
discussed above, $s > s_{\rm in} +1$ (e.g., Bicknell \& Begelman\
1996).

On the other hand, the very short synchrotron lifetime and the
extended character of the diffuse X-ray jet component discussed here
indicate that the appropriate acceleration process for the X-ray
emitting electrons cannot be connected exclusively with the localized
compact knots of the kpc-scale jet, but instead has to operate
effectively throughout the whole jet volume. Indeed, the maximum
propagation length of the jet electrons emitting 10\,keV synchrotron
photons within the 100\,$\mu$G jet (equipartition) magnetic field can
be estimated as $l_{\rm rad} = c \, \Gamma \, t_{\rm rad} \sim 10$\,pc
for $\Gamma \sim 2$, which clearly indicates the need for the
continuous acceleration of the radiating electrons in between  the
compact knots. Let us mention at this point, that the analysis of the
expected $\gamma$-ray emission from kpc-scale FR~I jets presented by
Stawarz et al.\ (2005) suggests that sub-equipartition magnetic
fields (which could increase $l_{\rm rad}$) are not likely in these
objects. The problem could disappear if many small scale {\it strong}
shocks are present (below the flux resolution limit for our point
source detection), possibly generated at the supersonic and highly
turbulent jet boundary layer. However, as argued in the previous
section, based on the constructed luminosity function for the resolved
jet-knots, the observed extended X-ray emission is most likely truly
diffuse in nature, and not just the pile-up of the unresolved compact
knots.

A physical process which can naturally account for the distributed
continuous acceleration of the radiating electrons within the whole
jet volume consists of interaction of the particles with turbulent MHD
waves. Indeed, by applying quasi-linear theory for particle-wave 
interactions (see, e.g., Schlickeiser 1989), one can find that with 
a reasonable set of physical parameters for the Cen~A jet, the available
maximum energies of the electrons accelerated in this way may be high 
enough to alow for efficient production of the synchrotron X-rays.
On the other hand, the resulting electron energy spectrum is expected to be
very flat: with a one-dimensional power spectrum of Alfv\'{e}nic turbulnce
$W(k)$\,$\propto$\,$k^{-q}$ and a `typical' range of turbulence spectral 
index $q \in (1, 2)$ --- usually obtained in phenomenological models for 
the astrophysical MHD turbulence (see, e.g., the recent review by Muller \& 
Biskamp\ 2003, and references therein) --- the steady-state solution for the
momentum diffusion equation is a power-law with 
$s_{\rm in} = q-1$ between 0 and 1
(see, e.g., Borovsky \& Eilek 1986 or Park \& Petrosian 1995).
However, the observed X-ray spectral index of the diffuse component of the 
Cen~A jet is steep, requiring an electron energy slope
of $s \simeq 3$. In particular, taking the expected injection spectrum
with $s_{\rm in}$ either 0 or 1, the strong (synchrotron) cooling
regime condition gives an observed electron energy index of $s = 2$ in
both cases (see Aharonian\ 2002), or an expected X-ray power-law
slope $\alpha_{\rm X} = 0.5$, i.e., much flatter than observed. It is
important to note, however, that the quasi-linear theory for
particle-wave interactions is of limited applicability
for physical conditions within extragalactic jets, which may be
highly turbulent, even with a possibly dynamically
important pressure due to ultra-relativistic particles at the jet
boundary (see Ostrowski 2000). In particular, the expected large ratio
of the escape to the radiative losses time scales within the jet (see 
Stawarz \& Ostrowski\ 2002) means that the stationary solution may
not be realizable due to accelerated particle back-reaction, or that the
turbulent diffusion could dominate the escape process. In such cases
the quasi-linear approach may no longer be valid. In addition, radiative 
effects within a non-uniform magnetic field can modify the high-energy 
part of the electron energy distribution in a non-trivial way. A set of 
such problems is under investigation.

\subsection{Hard Outer Sheath?}

Our analysis indicates the possible presence of hard X-ray
features within the outer (fragmentary) sheath of the Cen~A kpc-scale
jet, extending beyond the X-ray limb-brightened jet edges. The X-ray
spectral properties of the outer jet boundaries are most probably very
different from the ones characterizing the central jet flow. In
particular, the observed X-ray spectral index can be as flat
as $\alpha_{\rm X} < 0.5$. In addition, the physical connection of
these regions to the main flow is not clear, although they do roughly
coincide with the edges of the radio structure. One may note in this
context that detailed studies of the extragalactic radio jets using the
`spectral tomography' technique has also revealed the presence of some
features along the outer jet boundaries. In particular, Katz-Stone \&
Rudnick\ (1997) showed that the large-scale jet in the FR~I radio
galaxy 3C\,449, for which the radio spectral index $\alpha_{\rm R}
\approx 0.5$ is roughly constant along the whole jet, is surrounded by
a steep-spectrum radio `sheath'. Also, large-scale jets in the FR~II
radio galaxy Cygnus~A are placed in the edge-brightened `channels',
which are consistent with the enhancement of radio-emitting electrons
at the outer jet boundaries (Katz-Stone \& Rudnick\ 1994).

Unfortunately, the present X-ray data are not sufficient to
convincingly demonstrate the presence of a hard sheath around the
Cen~A jet. Let us however mention briefly a possible scenario which
can account for formation of such hard features, namely synchrotron
emission of the high-energy electrons which lose energy predominantly
due to inverse-Compton radiation in the Klein-Nishina regime
(hereafter `IC/KN' process; see Aharonian \& Ambartsumyan 1985;
Dermer \& Atoyan 2002; Kusunose \& Takahara 2005; Khangulian \&
Aharonian 2005; Moderski et al.\ 2005). An interesting aspect of this
scenario is that the observed electron spectrum can be flatter than
the injected one, with the asymptotic power-law slope $s \approx
s_{\rm in} - 1$. Let us therefore assume that the acceleration process
similar to the one operating within the 
jet acts also within
the outer boundaries of the outflow, where the magnetic field
intensity is possibly lower than within the jet. Thereby the
high-energy electrons cool partly due to the synchrotron emission,
producing X-ray photons, but mostly due to the IC/KN radiation at
$\gamma$-ray frequencies. The dominant IC/KN process thus flattens the
injected electron spectrum, leading to the observed flat X-ray
(synchrotron) continuum.

The Lorentz factor of electrons emitting 0.1\,keV synchrotron photons
is $\gamma_{\rm X} \approx 10^7 \, B_{-4}^{-1/2}$. Hence, the required
energy of the seed photons to be upscattered by these electrons in the
KN regime is $\varepsilon_0 > m_{\rm e} c^2 / 4 \, \gamma_{\rm X} \sim
10^{-2} \, B_{-4}^{1/2}$\,eV. At the same time, the energy density of
the seed photons has to be $U_0 > U_{\rm B} \approx 4 \times 10^{-10}
\, B_{-4}^2$\,erg\,cm$^{-3}$ in order to ensure the dominance of the
IC/KN losses. Taking the lowest expected value $B_{-4} \sim 0.01$,
this already excludes the cosmic microwave background radiation as the
source of seed photons. However, Cen~A is a giant elliptical galaxy
known for its strong infrared emission: $S_{\rm IR} = 23$\,Jy at
$\lambda_{\rm IR} = 12$\,$\mu$m (Knapp et al.\ 1990). Assuming that
all of this radiation is produced within $r = 1$\,kpc from the
galactic center (see Israel 1998), one can obtain a rough estimate for the
energy density of the galactic $\sim 0.1$\,eV photons at the position
of the kpc-scale Cen~A jet, $U_{\rm IR} = S_{\rm IR} \, d_{\rm L}^2 /
r^2 \, \lambda_{\rm IR} \sim 2 \times 10^{-12}$\,erg\,cm$^{-3}$, close
to the average value for giant ellipticals, $\sim
10^{-11}$\,erg\,cm$^{-3}$, found by Stawarz et al.\ (2003). Note that
this is about two orders of magnitude lower than the expected energy
density of the $\lambda_{\rm star} \sim 1$\,$\mu$m starlight energy
density (Stawarz et al.\ 2003). Also, the energy density of the
synchrotron infrared emission of the nuclear jet in Cen~A 
can be important in this respect. In particular, taking the observed
synchrotron luminosity of the Cen~A nuclear jet $L_{\rm bl} \approx 3
\times 10^{41}$\,erg\,s$^{-1}$ and its critical observed synchrotron
frequency $\nu_{\rm bl} \approx 10^{12}$\,Hz (Chiaberge et al.\ 2000),
one gets the appropriate energy density as measured by a stationary
observer located on the jet axis at $r = 1$\,kpc from the active
center $U_{\rm bl} = L_{\rm bl} \, \Gamma_{\rm bl}^3 / 4 \pi \, r^2 \,
c \sim 10^{-10}$\,erg\,cm$^{-3}$, and the appropriate critical
frequency $\Gamma_{\rm bl} \, \varepsilon_{\rm bl} \sim 0.03$\,eV, for
the expected bulk Lorentz factor of the nuclear jet $\Gamma_{\rm bl}
\sim 10$ (see Stawarz et al.\ 2003). Relative enhancement of synchrotron 
emission produced within a structured relativistic jet may constitute an 
alternative/additional source of the seed infrared photons (see in this 
context Ghisellini et al.\ 2005). Thus, we conclude that if only
the magnetic field in the outer sheath of the kpc-scale Cen~A jet is
$B \approx 10$\,$\mu$G can the suggested hard X-ray emission result
from synchrotron radiation of the high energy electrons characterized
by the standard initial spectrum $s_{\rm in} \simeq 2$, which cool
mainly by Comptonization of the nuclear jet emission (and also the
starlight photon field) in the KN regime.\footnote{We note that the
very high energy $\gamma$-ray flux of the outer sheath implied by
this model, $\sim 10^{-15}$ erg cm$^{-2}$ s$^{-1}$, is well below the
sensitivity limits of current Cherenkov Telescopes.}

\section{Conclusions}

We have presented an analysis of the diffuse X-ray emission from the
kpc-scale jet in Centaurus~A. We found that the extended emission is
most likely diffuse in nature rather than being the summation of unresolved
compact sources and that it exhibits a relatively flat-topped 
intensity profile with a 
steep, uniformly distributed spectral index of $\alpha_{\rm X} \approx
1$, similar to those observed in the compact knots. We interpret
the observed X-ray morphology in terms of a stratified jet model,
consisting of a relativistic outflow with an 
expanding
turbulent boundary shear layer. We argue that the observed X-ray
spectrum of this diffuse component, consistent with the synchrotron
emission produced in the strong cooling regime, requires continuous
acceleration of ultra-relativistic electrons within the whole jet
volume. We also note that the jet may be (partially) surrounded by an
outer sheath with a very hard X-ray spectrum. Both the presence and
origin of this feature are unclear: we suggest that it forms in a
turbulent region just outside the jet, separating the main outflow
from the external medium, where the conditions for the radiative
cooling processes are different from the ones characterizing the
jet. Although the present data do not allow us to
unequivocally demonstrate the presence of the hard outer sheath, or to
determine the appropriate particle acceleration mechanism acting
within (and possibly outside) the outflow, our analysis indicates that
(1) the jet structure is more complex than that visible on the
intensity maps alone, and (2) the acceleration and spectral evolution
of the radiating particles may be governed by many different processes
(including those such as magnetic field reconnection 
which we have not discussed), linked
to the complex magneto-hydrodynamical configuration of the outflow.

\acknowledgments

We thank an anonymous referee for his/her helpful comments which
helped clarify many of the issues presented in this paper. 
J.K.\ acknowledges support by the JSPS Kakenhi grant 14340061. 
\L .S.\ and M.O.\ were supported by the grant PBZ-KBN-054/P03/2001. \L .S.\ 
acknowledges also the ENIGMA Network through the grant HPRN-CT-2002-00321.

\begin{table}
\begin{center}
\caption{Newly detected X-ray knots and enhancements in the Cen~A jet}
\vspace{3mm}
\begin{tabular}{lrcccc}
\tableline\tableline
         & Distance  &  Angle & R.A. & Decl. & Net Counts \\
Knot ID  & (arcsec)  &  (deg) & (J2000) & (J2000) & (counts)   \\
\tableline
1 .......... &    9.96 ~& 36.65 & 13 25 28.32 & $-$43 01 02.8 & ~159$\pm$18~\\
2 .......... &   15.31 ~& 35.23 & 13 25 28.73 & $-$43 00 59.9 & 2682$\pm$55$^a$\\
3 .......... &   35.56 ~& 36.96 & 13 25 30.16 & $-$43 00 47.6 & ~244$\pm$17$^b$\\ 
4 .......... &   40.96 ~& 39.41 & 13 25 30.45 & $-$43 00 43.1 & ~100$\pm$12~\\
5 .......... &   55.65 ~& 35.06 & 13 25 31.69 & $-$43 00 37.2 & ~204$\pm$19~\\ 
6 .......... &   65.96 ~& 31.96 & 13 25 32.63 & $-$43 00 34.3 & ~123$\pm$13~\\
7 .......... &   80.78 ~& 40.48 & 13 25 33.12 & $-$43 00 17.0 & ~~42$\pm$~8~\\ 
8 .......... &   89.48 ~& 40.05 & 13 25 33.75 & $-$43 00 12.0 & ~~51$\pm$10~\\   
9 .......... &   86.64 ~& 37.35 & 13 25 33.79 & $-$43 00 16.9 & ~~53$\pm$10~\\
10 ..........&   82.92 ~& 32.40 & 13 25 33.89 & $-$43 00 24.9 & ~~44$\pm$~9~\\
11 ..........&  107.94 ~& 31.10 & 13 25 35.90 & $-$43 00 13.8 & ~~46$\pm$~9~\\
12 ..........&  163.57 ~& 38.42 & 13 25 39.10 & $-$42 59 28.6 & ~~16$\pm$~5~\\   
13 ..........&  209.89 ~& 37.59 & 13 25 42.52 & $-$42 59 02.6 & ~~38$\pm$~8~\\
\tableline
\tableline
\end{tabular}
\tablenotetext{a}{This extremely bright knot is only $\sim$ 1.5$''$ from
 the neighboring bright knot (RA=13:25:28.57, Dec=$-$43:01:00.1) and
 therefore was not previously able to be resolved from
 AX1 or knot2 in Table~2 of Kraft et al.\ 2003.}
\tablenotetext{b}{This jet-knot may be identified with the X-ray star
 J132530.1$-$430048 reported in Colbert et al.\ (2004).
}
\tablecomments{Thirteen new X-ray jet-knots and enhancements have been
 detected, in addition to those in Table~2 of Kraft et al.\ (2002), in
 the co-added, 0.4$-$8.0\,keV ACIS-S image. However, three knots in
 Kraft et al.\ (2002), namely NX1 (Knot-1), EX3 (Knot-19), and FX4
 (Knot-23), were not detected at their previously reported level for  
 various reasons. NX1 is only $\le$3$''$ apart from the nucleus and
 can hardly be detected in our 0.4$-$8.0 keV bandpass due to strong
 contamination of the bright nucleus emission. Kraft et al.\ (2002)
 claimed a detection of NX1 using only the 0.4$-$1.5 keV bandpass image, 
 since the heavily absorbed nucleus is expected to be suppressed. 
 This is also the case in our analysis, but we adopted a
 more conservative approach.
 We find no evidence for any enhancement in the jet emission
 around EX3, suggesting it may have been a statistical fluctuation.
 Finally, we suggest that FX4 is identical to FX3 reported in Kraft et
 al.\ (2002), but was regarded as a separate jet-knot due to 
 poor photon statistics. 
}
\end{center}
\end{table}

\begin{figure}[htb]
\begin{center}
\includegraphics[angle=0,scale=.5]{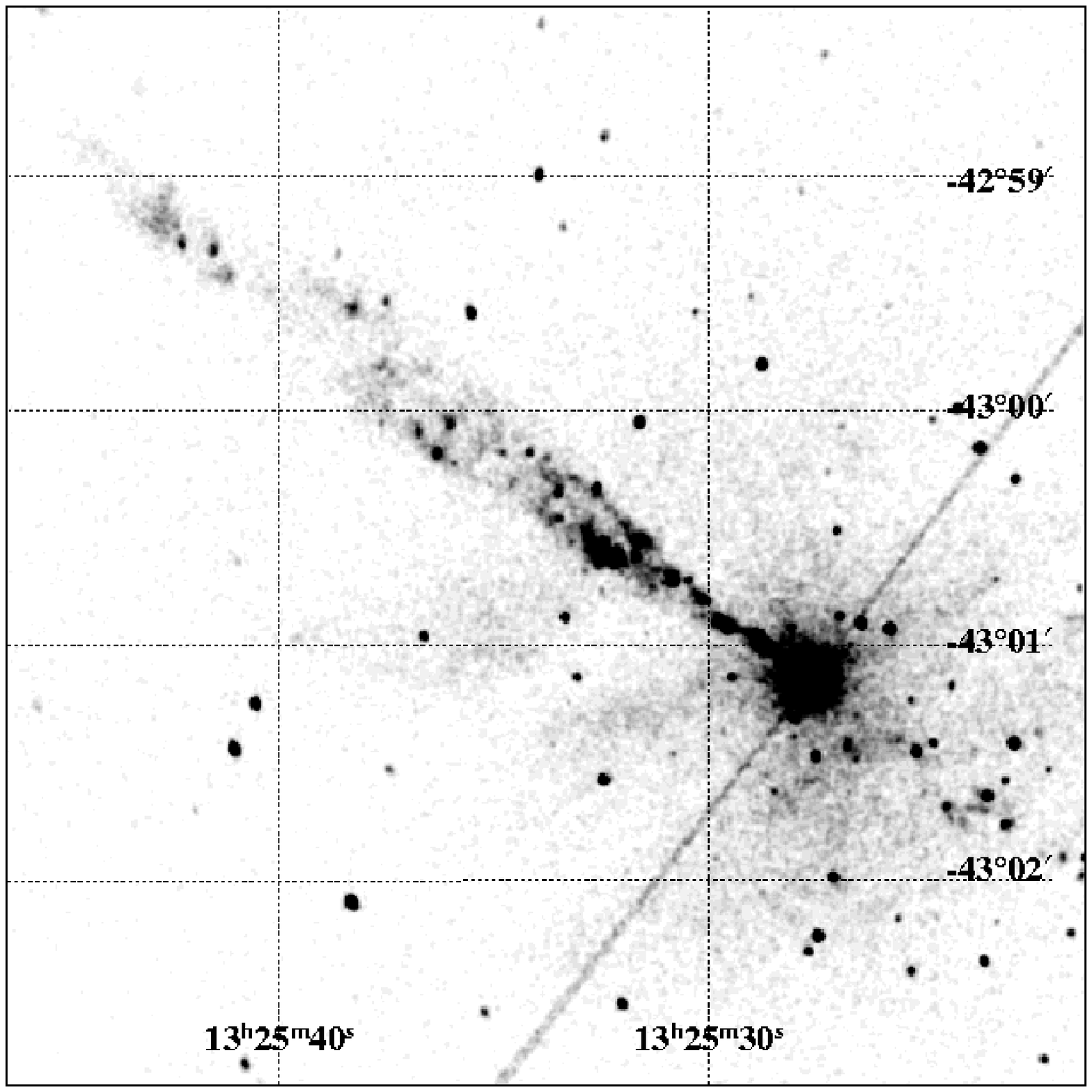}
\includegraphics[angle=0,scale=.5]{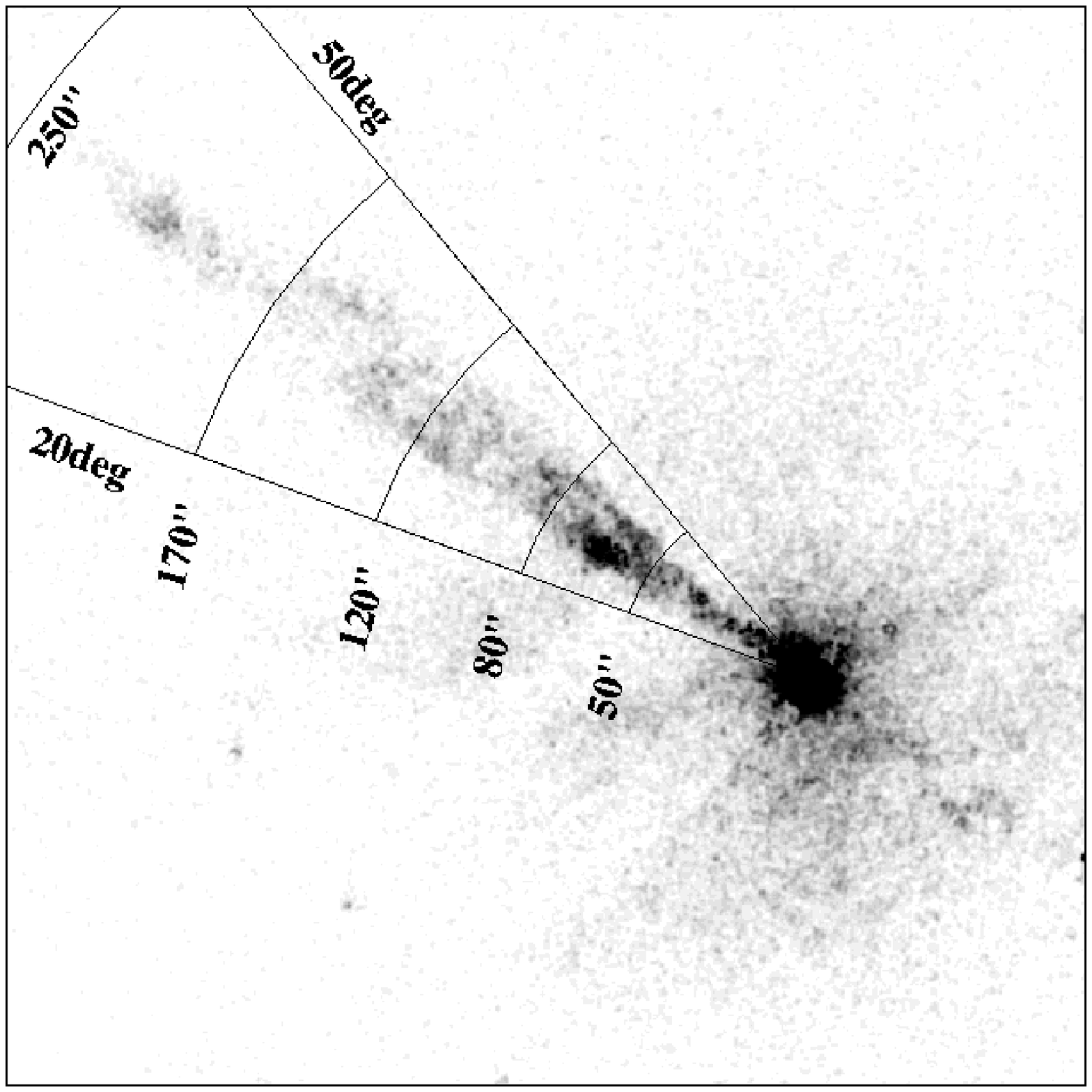}
\caption{{\it top:} The 0.4$-$8.0\,keV X-ray image of the Cen~A jet. 
Archival data from Obs ID 2978 and 3965 are combined. 
The image is smoothed with a $\sigma$\,$=$\,0.5 arcsec Gaussian. 
{\it bottom:} The 0.4$-$8.0\,keV X-ray image of the Cen~A jet, 
after subtracting the point sources and filling in the blanks 
with values interpolated from surrounding background regions. 
More details are given in the text. 
Definitions of the angle $\Theta$ and radial distance $r$ 
are given in the panel.}
\end{center}
\end{figure}

\begin{figure}[htb]
\begin{center}
\includegraphics[angle=90,scale=0.7]{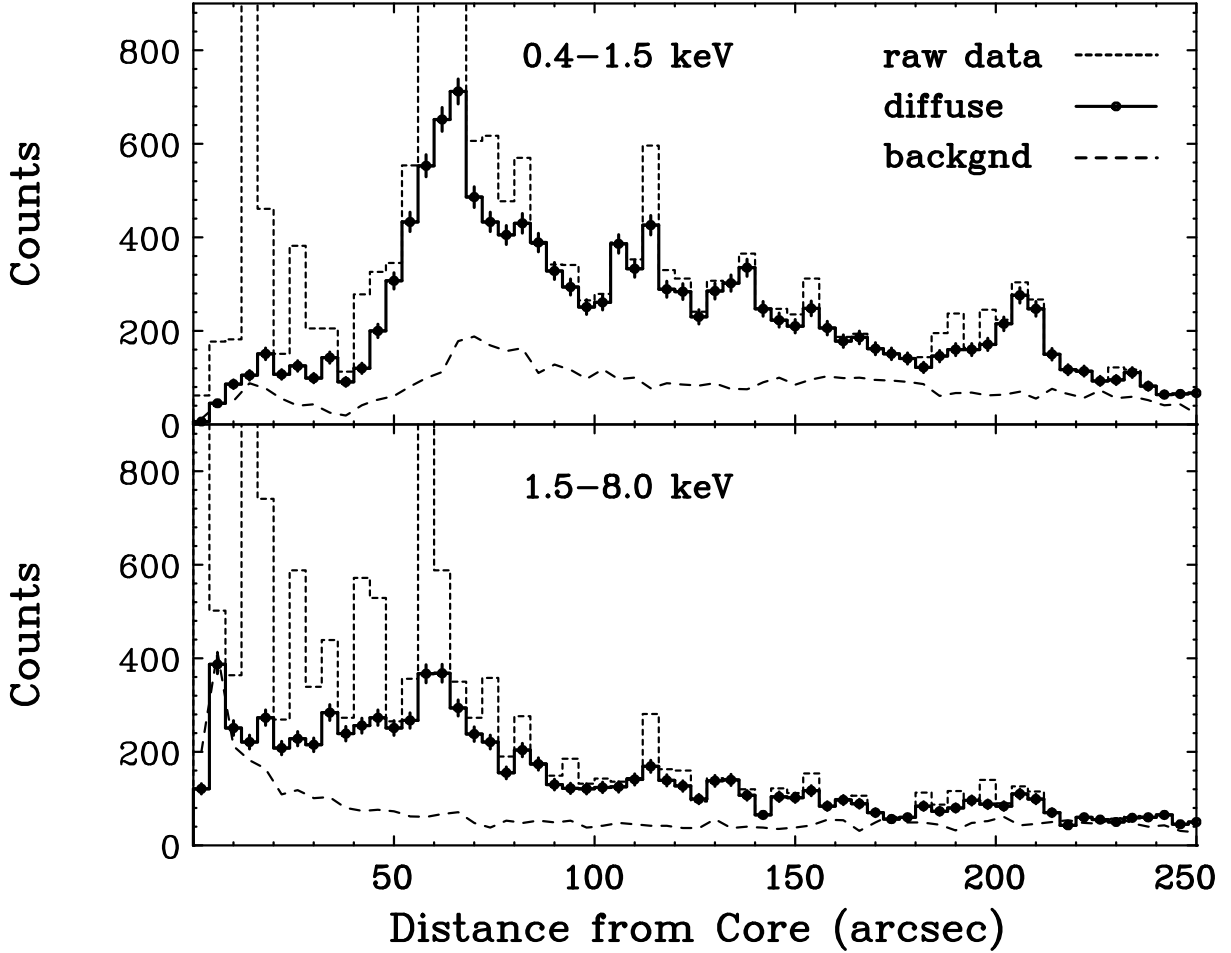}
\caption{Longitudinal jet profile of the Cen~A 
in the soft X-ray energy band (0.4--1.5\,keV; {\it upper}) 
and the hard X-ray energy band (1.5--8.0\,keV; {\it lower}), 
where the photons counts are integrated over the position angles 
25$\deg$ $\le$ $\Theta$ $\le$ 45$\deg$. 
The dotted histogram shows the profile derived from the raw X-ray image 
(Figure~1 {\it top}), whereas the solid histogram with data points 
shows the one obtained for the diffuse emission only 
(Figure~1 {\it bottom}). The dashed line shows the background 
profile compiled from 15$\deg \le \Theta \le$\,25$\deg$ 
and 45$\deg \le \Theta \le$\,55$\deg$ regions.}
\end{center}
\end{figure}

\begin{figure}[htb]
\begin{center}
\includegraphics[angle=90,scale=0.7]{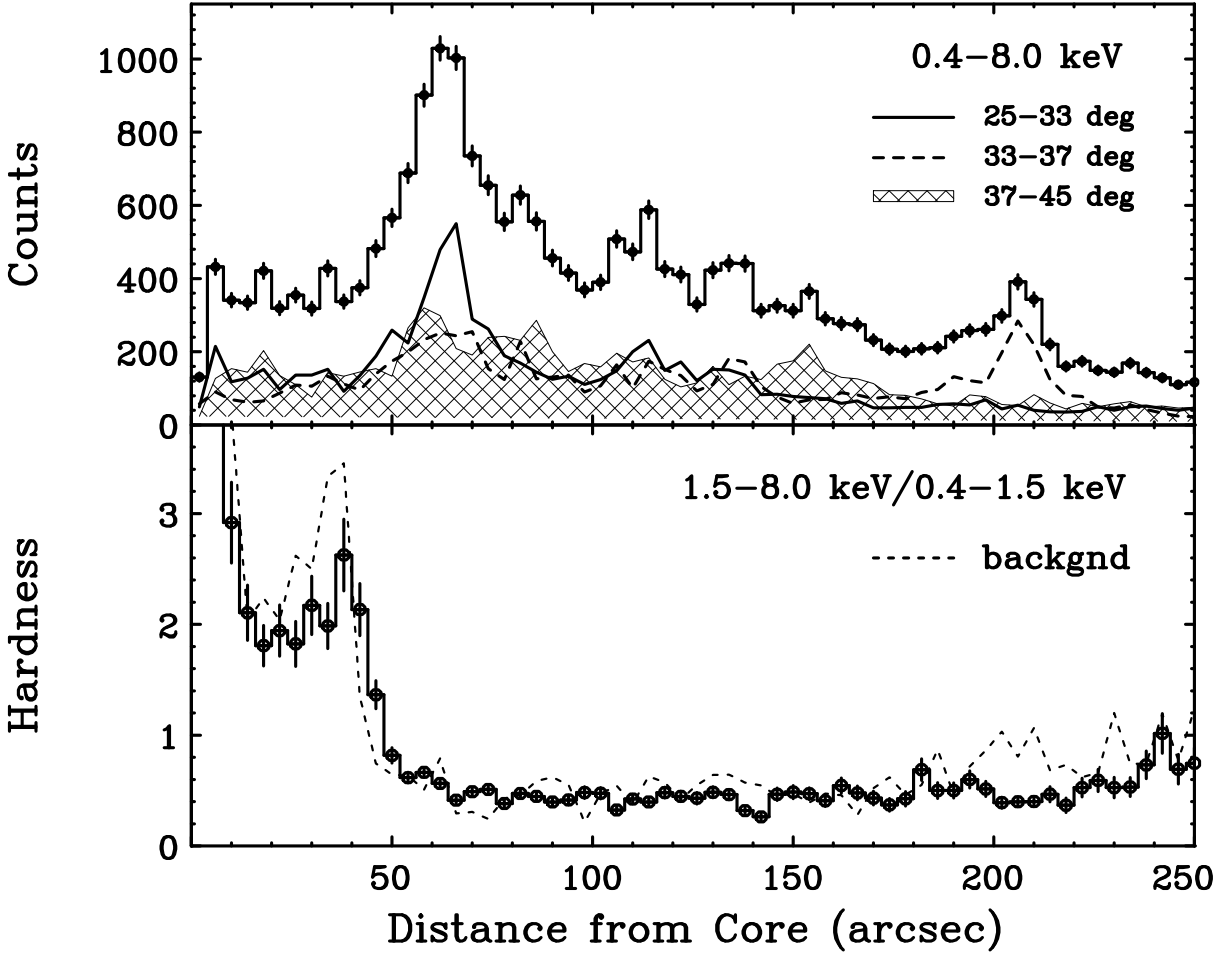}
\caption{{\it Upper:} The observed longitudinal X-ray intensity
 profile of the diffuse jet component in the 0.4$-$8.0\,keV photon
 energy band.  Photon counts are integrated over the position angle
 ranges 25$\deg \le \Theta \le$\,45$\deg$ (histogram), 
 25$\deg \le \Theta \le$\,33$\deg$ (solid line), 
 33$\deg \le \Theta \le$\,37$\deg$ (dashes), and 
 37$\deg \le \Theta \le$\,45$\deg$ (hatched).
 {\it Lower:} Variation of the hardness ratio
 along the main jet axis.  Hardness is defined as the ratio between
 the X-ray counts measured in the soft X-ray energy band
 (0.4$-$1.5\,keV) and the hard X-ray energy band (1.5$-$8.0\,keV).
 Note that the hardness ratio changes significantly at the distance
 $r_{\rm c} \simeq$\,50$''$, but stays almost constant at larger
 distances.  The dotted line shows the hardness changes for the
 background compiled from 15$\deg \le \Theta \le$\,25$\deg$ and
 45$\deg \le \Theta \le$\,55$\deg$ regions.}
\end{center}
\end{figure} 

\begin{figure}[htb]
\begin{center}
\includegraphics[angle=90,scale=0.7]{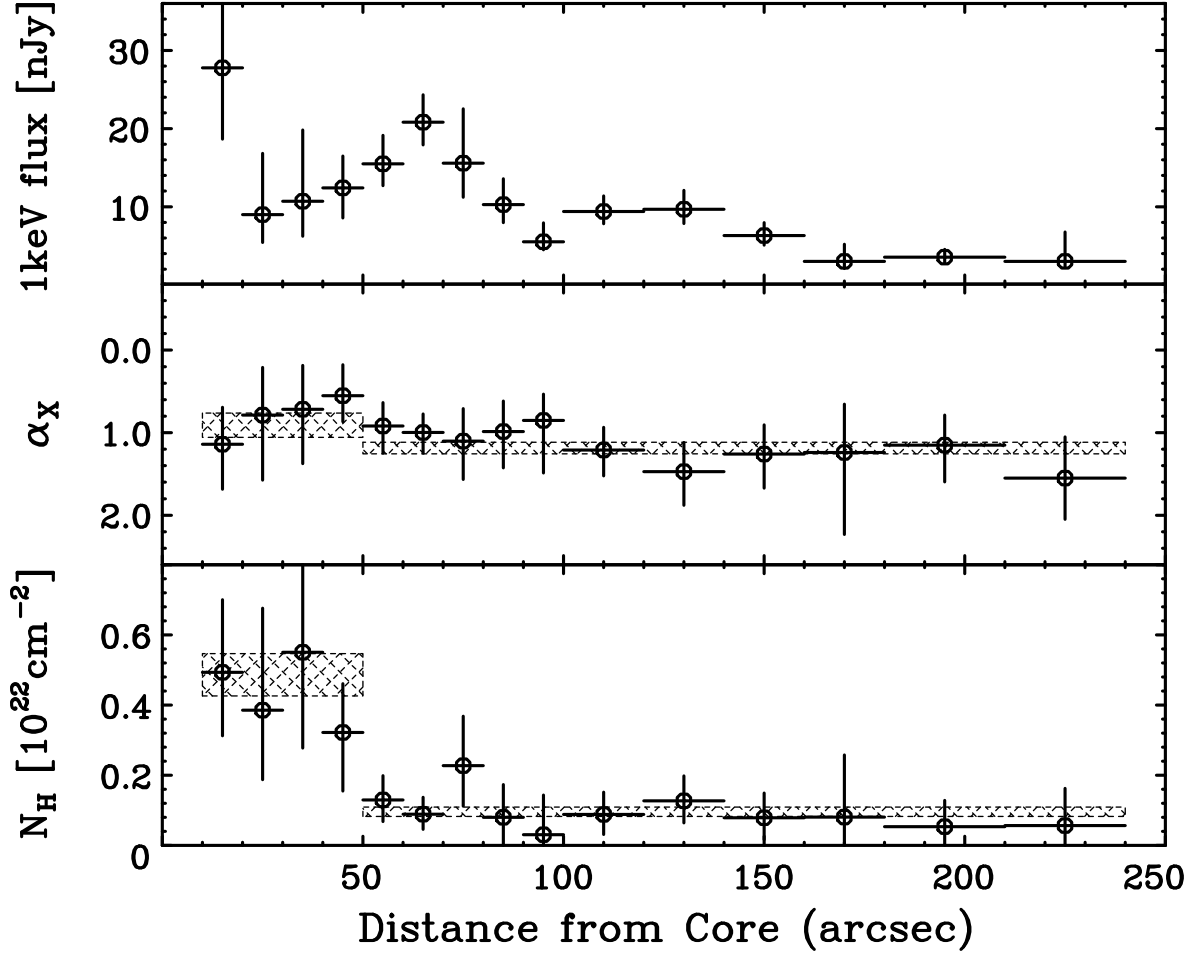} 
\caption{Variation of the X-ray flux density measured at 1\,keV ({\it top}),
 the X-ray energy spectral index $\alpha_{\rm X}$ ({\it middle}), and
 the absorbing column density $N_{\rm H}$ ({\it bottom}) of the
 diffuse emission along the jet.  Note that $N_{\rm H}$ changes
 significantly at $r_{\rm c} \simeq$\,50$''$, but that the energy
 spectral index is almost constant ($\alpha_{\rm X}$\,$\simeq$\,1)
 along the jet.  Hatched regions show the best-fit parameters and
 their 1\,$\sigma$ uncertainties for the combined spectral fittings
 within 10$'' \le r \le$\,50$''$ and 50$'' \le r \le$\,250$''$, respectively.}
\end{center}
\end{figure} 

\begin{figure}[htb]
\begin{center}
\includegraphics[angle=90,scale=0.5]{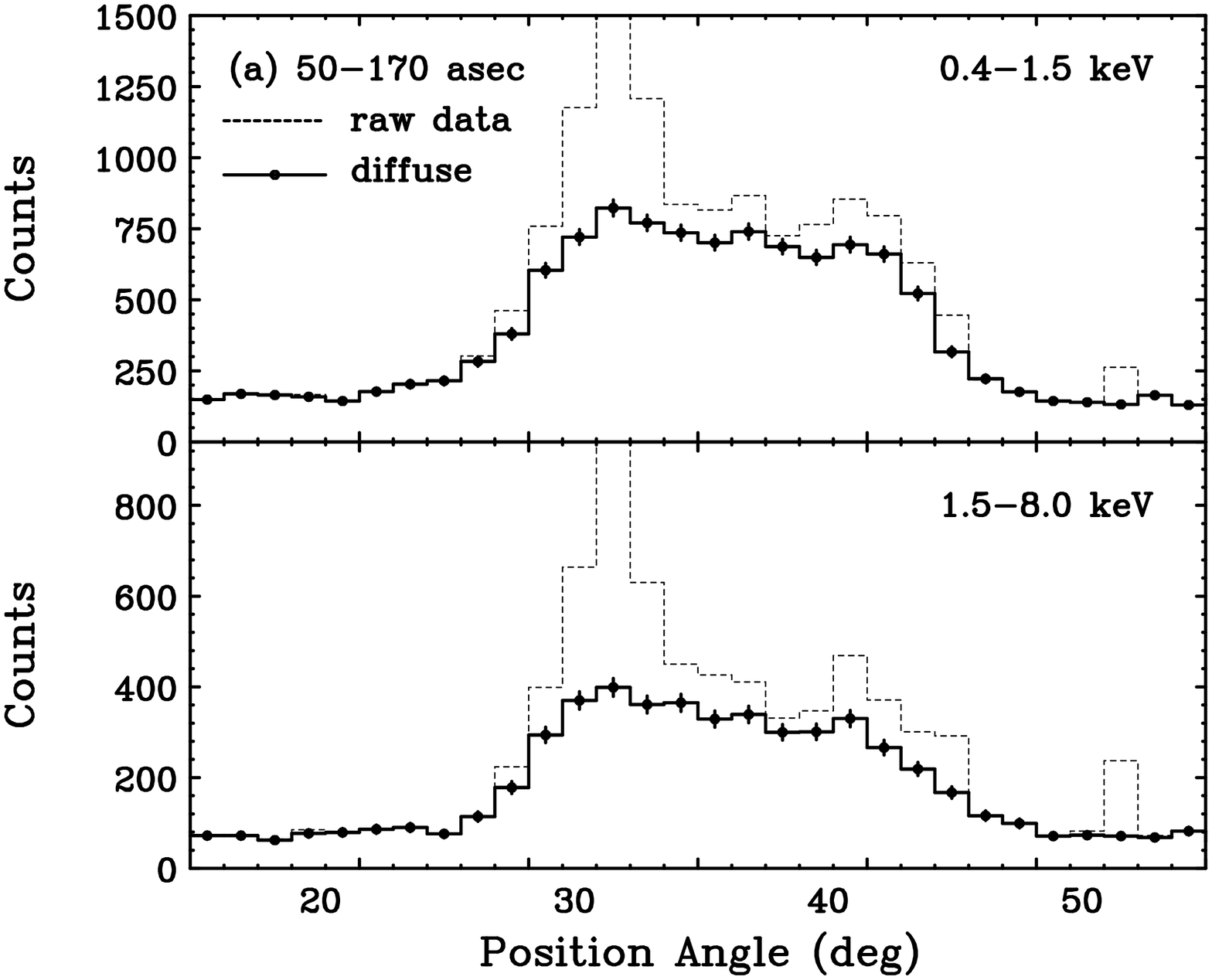}
\includegraphics[angle=90,scale=0.5]{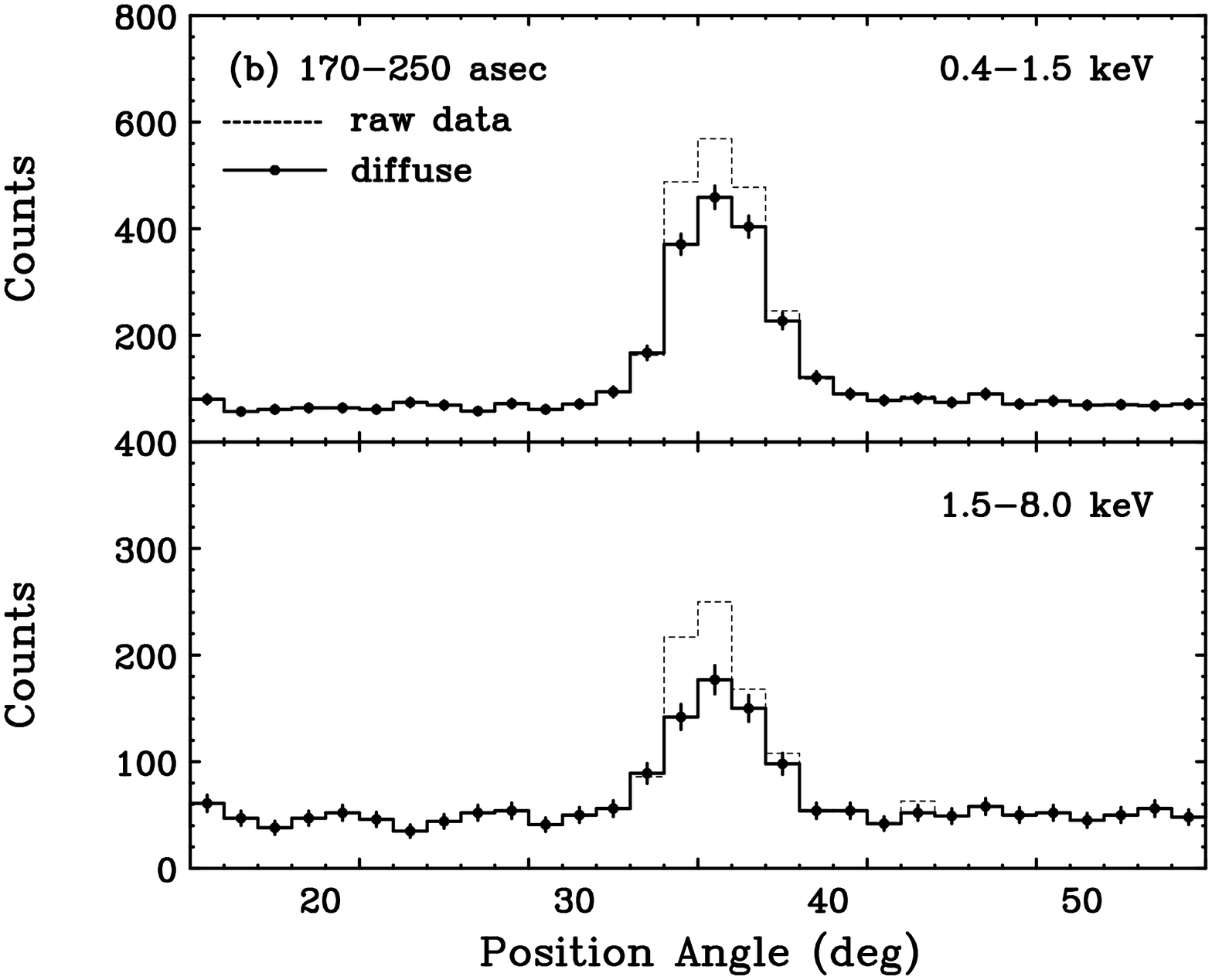}
\caption{Transverse jet profile of the Cen~A in the soft X-ray energy
 band (0.4--1.5\,keV) and the hard X-ray energy band (1.5--8.0\,keV).
 The profiles for the inner jet region ({\it upper}; 
 (a) 50$'' \le r \le$\,170$''$) and the outer jet regions ({\it lower}; 
 (b) 170$'' \le r \le$\,250$''$) are given in separate panels. The dashed
 histogram shows the profile derived from the raw X-ray image
 (Figure~1 {\it upper}), whereas the solid histogram with the data
 points shows that obtained for the diffuse emission only (Figure~1
 {\it bottom}).}
\end{center}
\end{figure}

\begin{figure}[htb]
\begin{center}
\includegraphics[angle=90,scale=0.35]{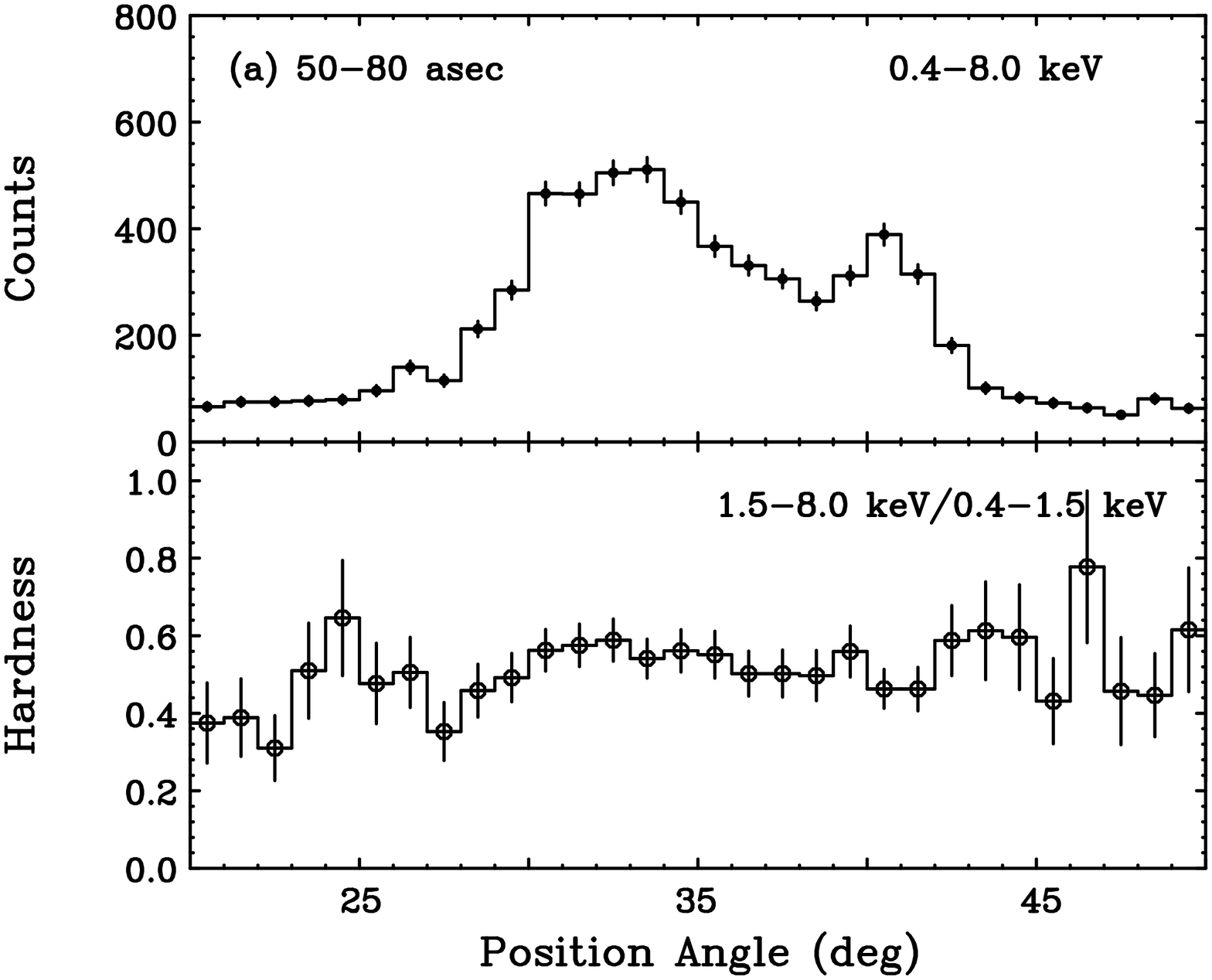}
\includegraphics[angle=90,scale=0.35]{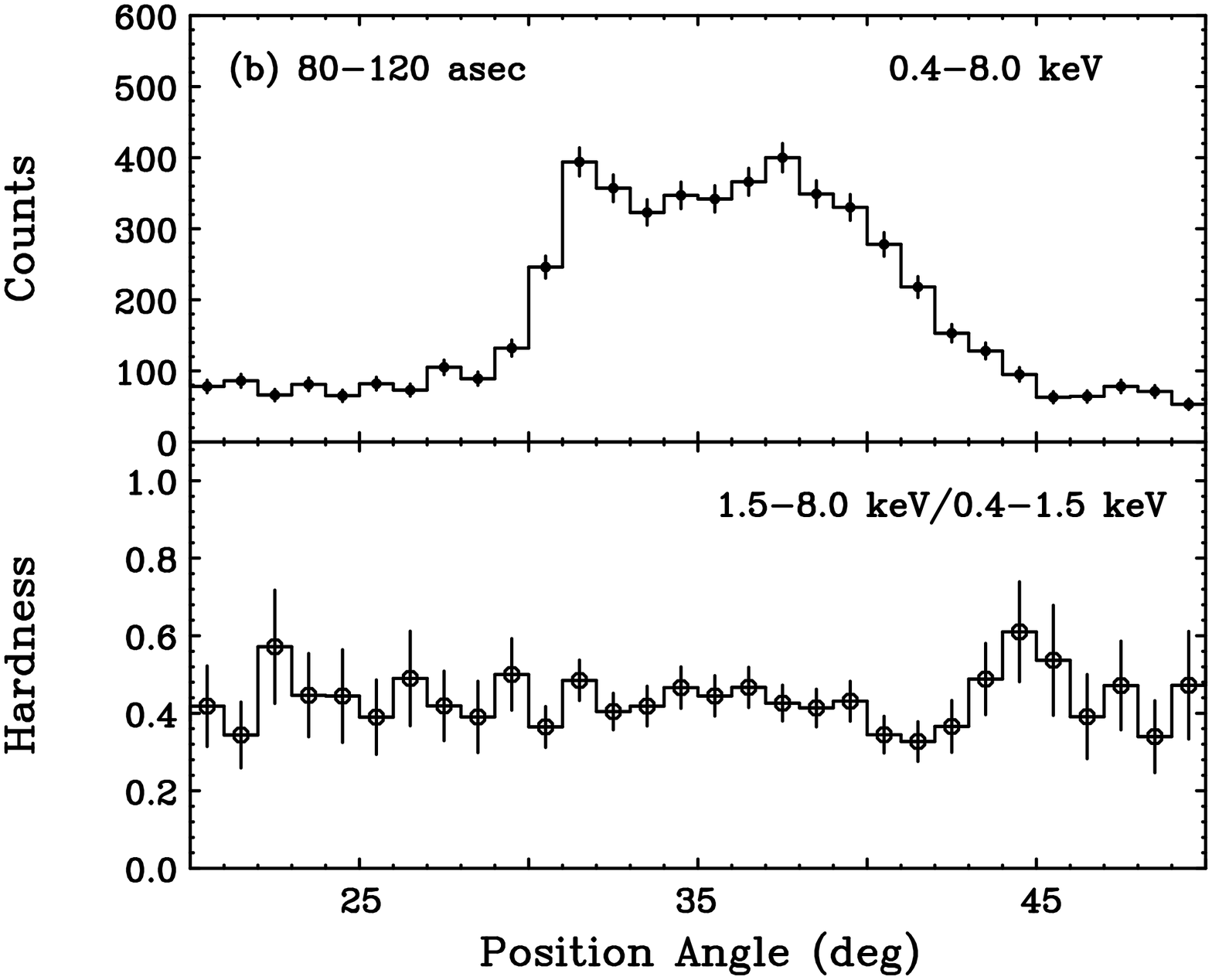}
\includegraphics[angle=90,scale=0.35]{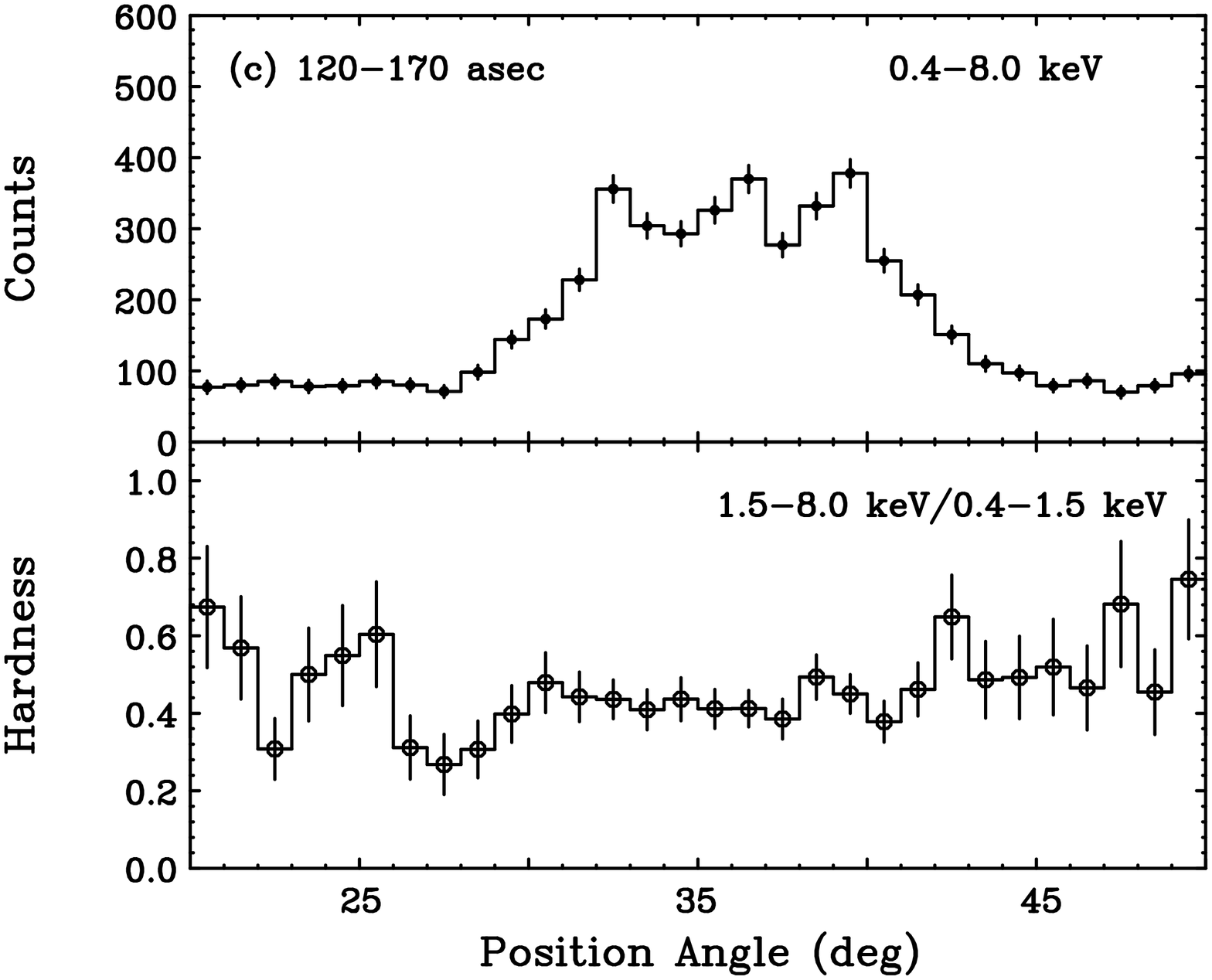}
\includegraphics[angle=90,scale=0.35]{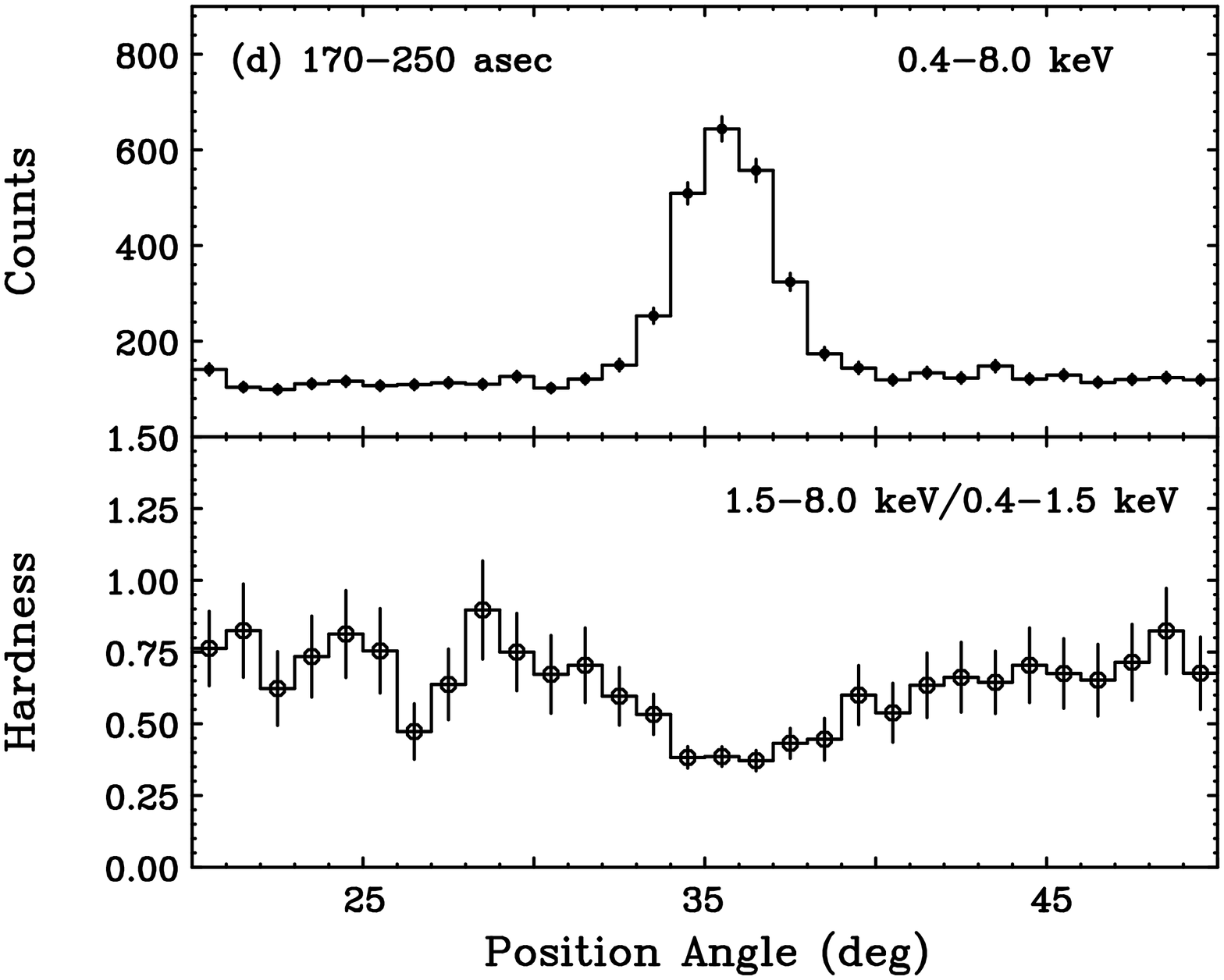}
\caption{{\it Upper:} The observed transverse X-ray intensity profile
 of the diffuse jet component in the 0.4$-$8.0\,keV photon energy
 band. {\it Lower:} Variation of the hardness ratio in the direction
 perpendicular to the jet main axis. Hardness is defined as the ratio
 between the X-ray counts measured in the soft X-ray energy band
 (0.4$-$1.5\,keV) and the hard X-ray energy band (1.5$-$8.0\,keV).
 The profiles are given in separate panels for various distances from
 the nucleus, 
 (a) 50$'' \le r \le$\,80$''$, (b) 80$'' \le r \le$\,120$''$, 
 (c) 120$'' \le r \le$\,170$''$, and (d) 170$'' \le r \le$\,250$''$. 
 Note the possible hardness variations near
 the edges of the jet, $\Theta \sim$\,25$\deg$ and $\Theta \sim$\,45$\deg$.}
\end{center}
\end{figure} 

\begin{figure}[htb]
\begin{center}
\includegraphics[angle=90,scale=0.35]{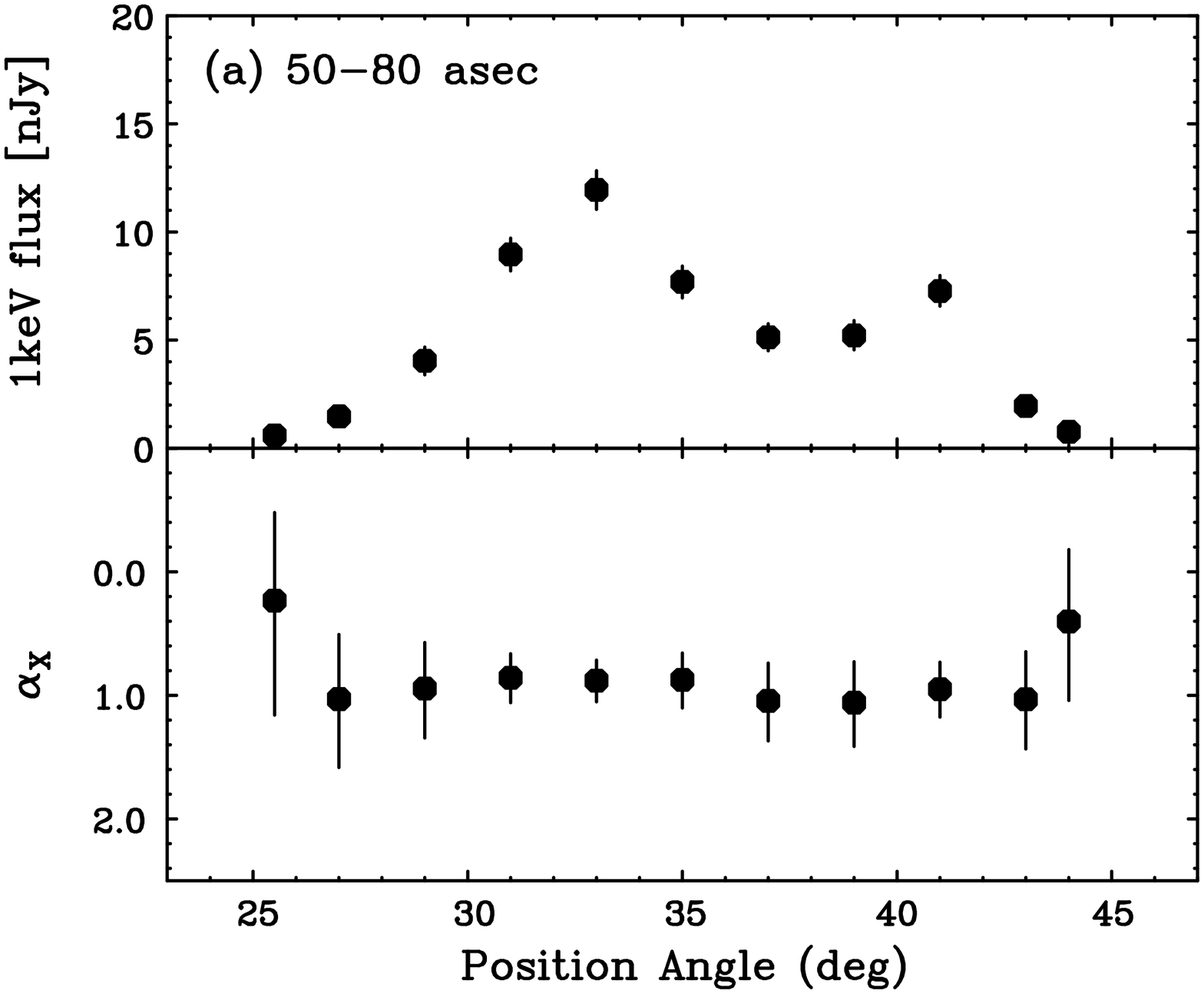} 
\includegraphics[angle=90,scale=0.35]{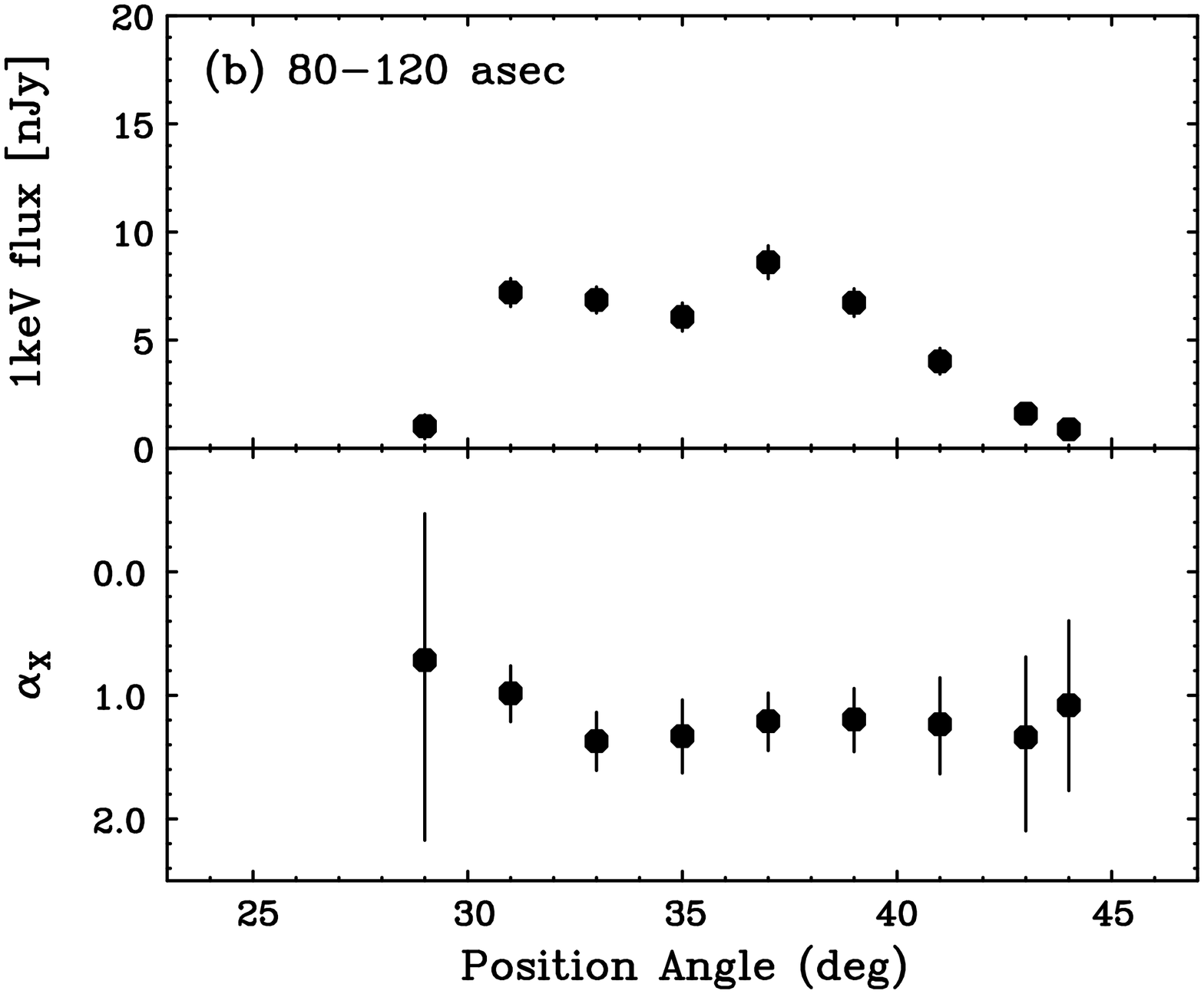}
\includegraphics[angle=90,scale=0.35]{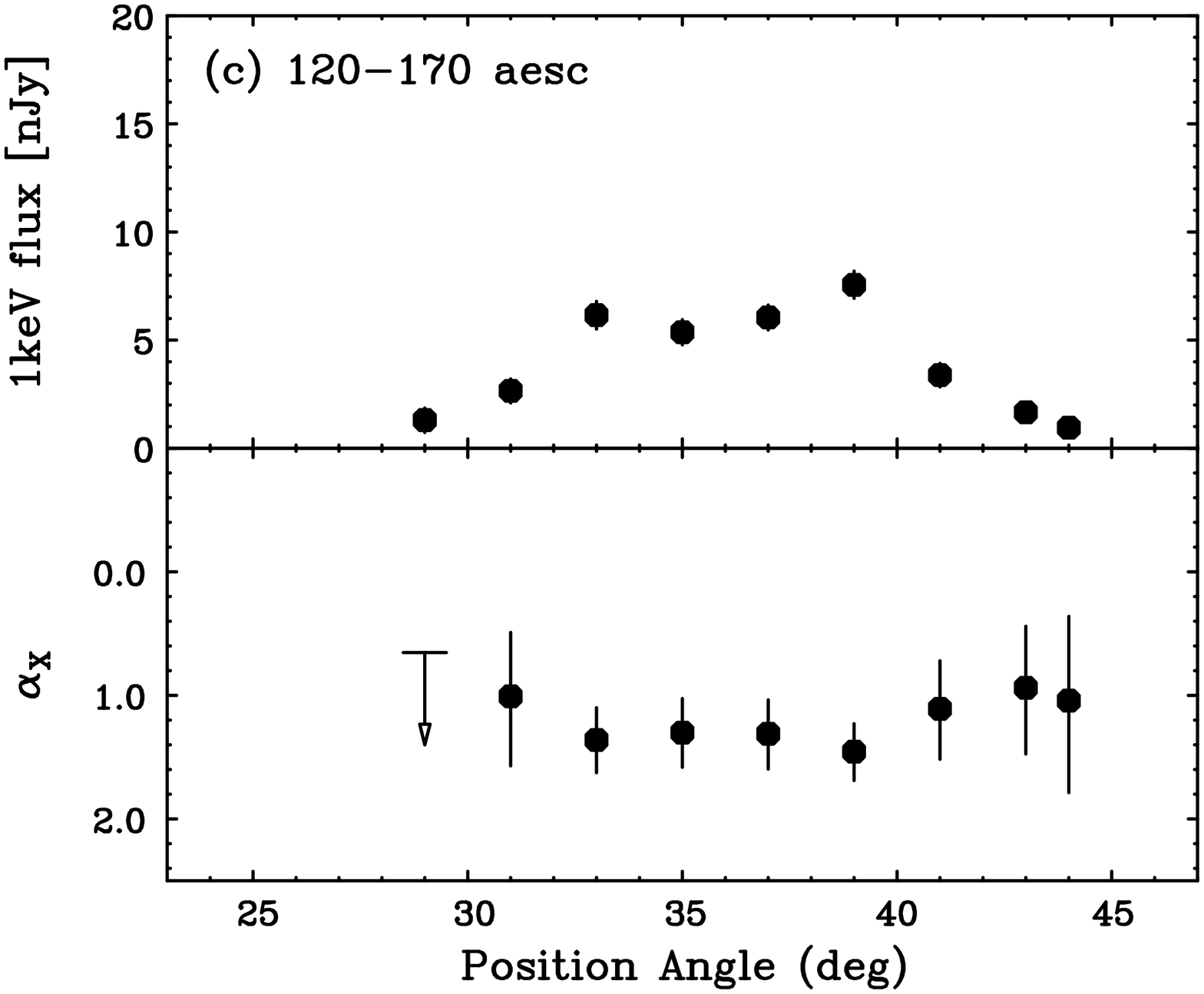}
\includegraphics[angle=90,scale=0.35]{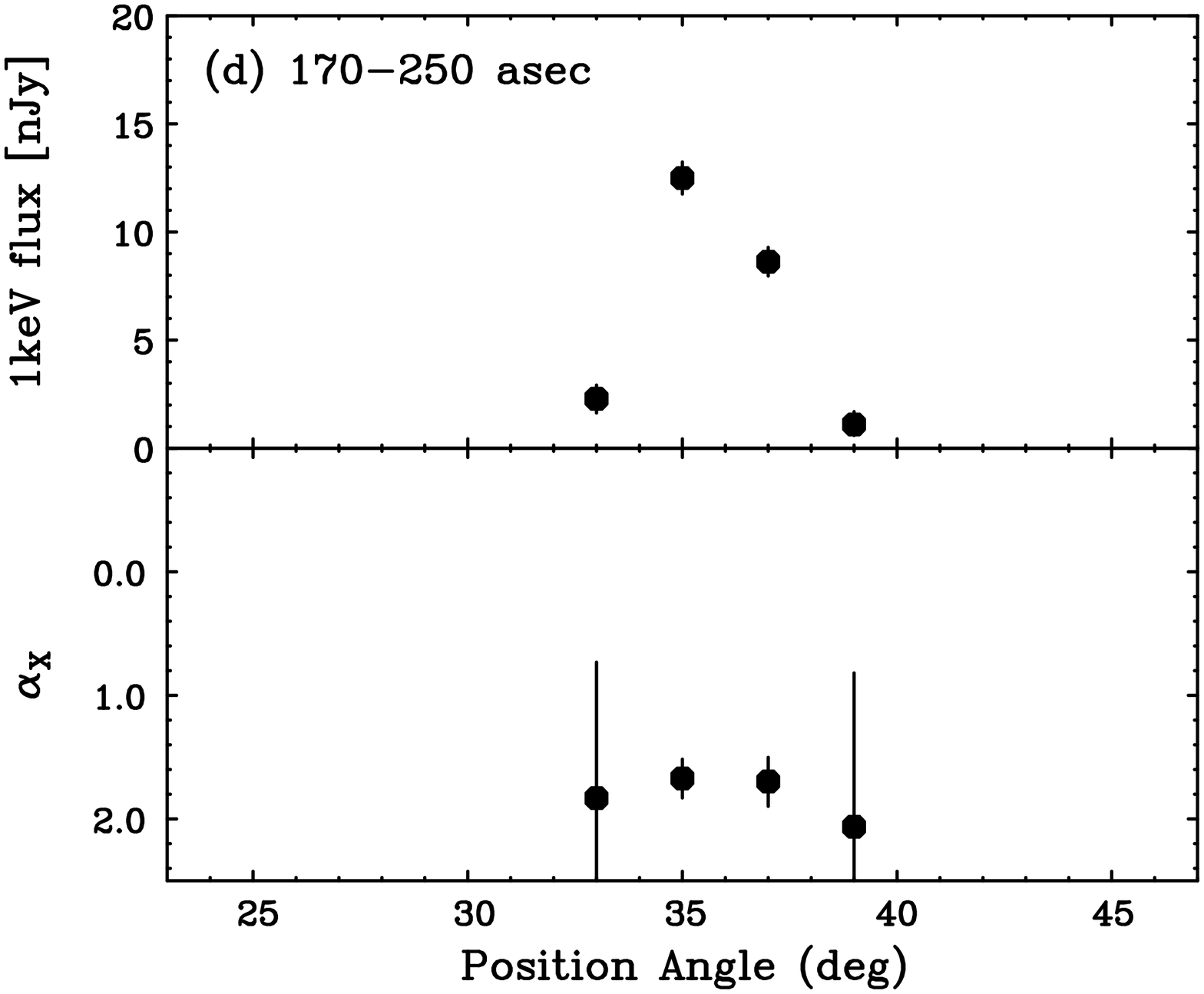}   
\caption{Variation of the X-ray intensity measured at 1\,keV 
 ({\it top}), and the energy spectral index ({\it bottom}) across the
 jet, in $\Delta \Theta$\,$=$\,2$\deg$ steps.  The contribution of
 point sources is simply removed, and {\it dmfilth} is not used in
 order to avoid any possible artifacts in the spectral analysis.
 Reduction of integration regions due to the point source subtraction
 is corrected in estimating the 1\,keV flux.  The fitting model
 consists of a power-law function and the assumed Galactic absorption
 column density $N_{\rm H}$\,$=$\,0.96\,$\times$\,10$^{21}$\,cm$^{-2}$. }
\end{center}
\end{figure}

\begin{figure}[htb]
\begin{center}
\includegraphics[angle=90,scale=0.7]{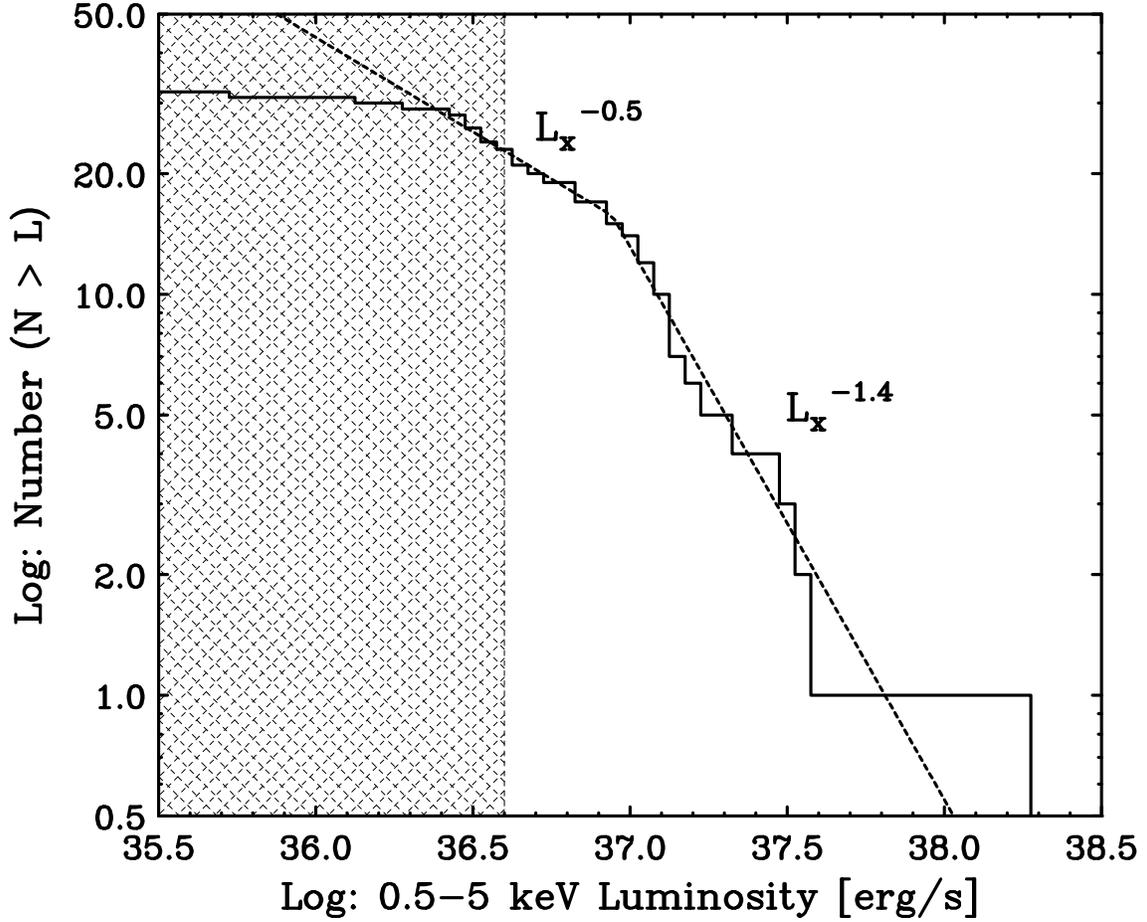} 
\caption{Luminosity function of the knots in the jet, constructed
 excluding the knots of the innermost part of the outflow 
 (0$'' \le r \le$\,50$''$).  The X-ray luminosity is the absorption
 corrected luminosity in the 0.5$-$5.0\,keV band, assuming a
 power-law spectrum with energy spectral index $\alpha_{\rm X}$\,=\,1.0
 and $N_{\rm H}$ = 0.96$\times$10$^{21}$\,cm$^{-2}$. The best fit
 broken power-law model, excluding the hatched region where our
 sample is incomplete and biased 
 ($L_{\rm X} \le$4$\times$10$^{36}$\,erg s$^{-1}$), is also shown.}
\end{center}
\end{figure} 

\end{document}